\documentclass[a4paper,aps,pre,floatfix,superscriptaddress,nofootinbib,showpacs,twocolumn]{revtex4-1}

\usepackage[pdftex]{graphicx}
\usepackage{hyperref}

\usepackage{amsmath}
\usepackage{amsfonts}
\usepackage{amssymb}
\usepackage{pifont}
\usepackage[english]{babel}

\usepackage[pdftex,usenames]{color}

\newcommand{\av}[1]{\langle {#1} \rangle}

\begin{document}

\title{ Random walks on temporal networks}

\author{Michele Starnini} 

\address{Departament de F\'\i sica i Enginyeria Nuclear,
  Universitat Polit\`ecnica de Catalunya, Campus Nord B4, 08034
  Barcelona, Spain}

\author{Andrea Baronchelli} 

\address{Department of Physics, College of Computer and
  Information Sciences, Bouv\'e College of Health Sciences,
  Northeastern University, Boston MA02120, USA}
  
\author{Alain Barrat}

\address{Centre de Physique Th\'eorique, Aix-Marseille Univ, CNRS
  UMR 7332, Univ Sud Toulon Var, 13288 Marseille cedex 9, France}
\address{Data Science Laboratory, ISI Foundation, Torino, Italy}

\author{Romualdo Pastor-Satorras} 
\address{Departament de F\'\i sica i Enginyeria Nuclear,
  Universitat Polit\`ecnica de Catalunya, Campus Nord B4, 08034
  Barcelona, Spain}

\date{\today}

\begin{abstract}
  Many natural and artificial networks evolve in time. Nodes and
  connections appear and disappear at various timescales, and their
  dynamics has profound consequences for any processes in which they
  are involved. The first empirical analysis of the temporal patterns
  characterizing dynamic networks are still recent, so that many
  questions remain open. Here, we study how random walks, as
  paradigm of dynamical processes, unfold on temporally evolving
  networks. To this aim, we use empirical dynamical networks of
  contacts between individuals, and characterize the fundamental
  quantities that impact any general process taking place upon
  them. Furthermore, we introduce different randomizing strategies
  that allow us to single out the role of the different properties of
  the empirical networks. We show that the random walk exploration is
  slower on temporal networks than it is on the aggregate projected
  network, even when the time is properly rescaled. In particular, we
  point out that a fundamental role is played by the temporal
  correlations between consecutive contacts present in the
  data. Finally, we address the consequences of the intrinsically
  limited duration of many real world dynamical networks. Considering
  the fundamental prototypical role of the random walk process, we
  believe that these results could help
  to shed light on the behavior of more complex dynamics on
  temporally evolving networks.
\end{abstract}

\pacs{05.40.Fb, 89.75.Hc, 89.75.-k}

\maketitle

\section{Introduction}

Many real networks are dynamic structures in which connections appear,
disappear, or are rewired on various timescales \cite{Holme:2011fk}.
For example, the links representing social relationships in social
networks \cite{wass94} are a static representation of a succession of
contact or communication events, which are
constantly created or terminated between pairs of individuals
(actors). Such temporal evolution is an intrinsic feature of many
natural and artificial networks, and can have profound consequences
for the dynamical processes taking place upon them. Until recently
however, a large majority of studies about complex networks have
focused on a static or aggregated representation, in which all the
links that appeared at least once coexist. This is the case, for
example, in the seminal works on scientific collaboration networks
\cite{newmancitations01}, or on movie costarring networks
\cite{Barabasi:1999}. In particular, dynamical processes have mainly
been studied on static complex networks \cite{BBV}.

In recent years, the interest towards the temporal dimension of the
network description has blossomed. Empirical analyses have revealed
rich and complex patterns of dynamic evolution
\cite{Hui:2005,PhysRevE.71.046119,Onnela:2007,Gautreau:2009,10.1371/journal.pone.0011596,Tang:2010,Bajardi:2011,Stehle:2011nx,Miritello:2011,Karsai:2011,Holme:2011fk},
pointing out the need to characterize and model them
\cite{Scherrer:2008,Hill:2009,Gautreau:2009,PhysRevE.81.035101,PhysRevE.83.056109}.
At the same time, researchers have started to study how the temporal evolution of the network substrate impacts 
the behavior of dynamical processes such as 
epidemic spreading \cite{Rocha:2010,Isella:2011,Stehle:2011nx,Karsai:2011,Miritello:2011,dynnetkaski2011},
synchronization \cite{albert2011sync}, percolation \cite{Parshani:2010,Bajardi:2011}
and social consensus \cite{consensus_temporal_nrets_2012}.

Here, we focus on the dynamics of a random walker
exploring a temporal network \cite{WeissRandomWalk,hughes,lovasz}. The
random walk is indeed the simplest diffusion model, and its dynamics
provides fundamental hints to understand the whole class of diffusive
processes on networks. Moreover, it has relevant applications in
such contexts as spreading dynamics (i.e. virus or opinion spreading)
and searching.  For instance, assuming that each vertex knows only
about the information stored in each of its nearest neighbors, the
most naive economical strategy is the random walk search, in which the
source vertex sends one message to a randomly selected nearest
neighbor \cite{PhysRevE.64.046135,Lv:2002,BBV}. If that vertex has the
information requested, it retrieves it; otherwise, it sends a message
to one of its nearest neighbors, until the message arrives to its
finally target destination. Thus, the random walk represents a lower
bound on the effects of searching in the absence of any information in
the network, apart form the purely local information about the
contacts at a given instant of time.

In our study, we consider as typical examples of temporal networks the
dynamical sequences of contact between individuals in various social
contexts, as recorded by the SocioPatterns project
\cite{Sociopatternsweb,10.1371/journal.pone.0011596}. These datasets
contain indeed the time-resolved patterns of face-to-face co-presence
of individuals in settings such as conferences, with high temporal
resolution: for each contact between individuals, the starting and
ending times are registered by the measuring infrastructure, giving
access to the timing {\it and} duration of contacts.

The paper is structured as follows. In Sec \ref{sec:overv-rand-walks}
we review some of the fundamental results for random walks on static
networks. In Sec. \ref{sec:real-dynam-netw} we describe the empirical dynamical
networks considered: we recall some basic definitions,
present an analysis of the datasets, and introduce suitable
randomization procedures, which will help later on to pinpoint the
role of the correlations in the real data. In
Sec. \ref{sec:rand-walks-empir} we write down mean-field equations for
the case of maximally randomized dynamical contact networks, and in
Sec. \ref{sec:numerical-simulations} we investigate the random walk
dynamics numerically, focusing on the exploration properties and on
the mean first passage times. Sec. \ref{sec:random-walks-finite} is
devoted to the analysis of the impact of the finite temporal duration
of real time series. Finally, we summarize our results and comment on
some perspectives in Sec \ref{sec:disc-concl}.

\section{A short overview of random walks on static networks}
\label{sec:overv-rand-walks}

The random walk (RW) process is defined by a
walker that, located on a given vertex $i$ at time $t$, hops to a
nearest neighbor vertex $j$ at time $t+1$.

In binary networks, defined by the adjacency matrix $a_{ij}$ such that
$a_{ij}=1$ is $j$ is a neighbor of $i$, and $a_{ij}=0$ else, the
transition probability at each time step from $i$ to $j$ is
\begin{equation}
  \label{eq:1}
  p_b(i \to j) = \frac{a_{ij}}{\sum_r a_{ir}} \equiv \frac{a_{ij}}{k_i},
\end{equation}
where $k_i= \sum_j a_{ij}$ is the degree of vertex $i$: the walker
hops to a nearest neighbor of $i$, chosen uniformly at random among
the $k_i$ neighbors, hence with probability $1/k_i$ (note that we
consider here undirected networks with $a_{ij}=a_{ji}$, but the
process can be considered as well on directed networks).  In
weighted networks with a weight matrix $\omega_{ij}$, the transition
probability takes instead the form
\begin{equation}
  \label{eq:2}
  p_w(i \to j) = \frac{w_{ij}}{\sum_r w_{ir}} \equiv \frac{w_{ij}}{s_i},
\end{equation}
where $s_i=\sum_j \omega_{ij}$ is the strength of vertex $i$
\cite{Barrat16032004}. Here the walker chooses a nearest neighbor with
probability proportional to the weight of the corresponding connecting
edge.

The basic quantity characterizing random walks in networks is the
\textit{occupation probability} $\rho_i$, defined as the steady state
probability (i.e., measured in the infinite time limit) that the
walker occupies the vertex $i$, or in other words, the steady state
probability that the walker will land on vertex $i$ after a jump from
any other vertex. Following rigorous master equation arguments, it is
possible to show that the occupation probability takes the form
\cite{PhysRevLett.92.118701,wu07:_walks}
\begin{equation}
  \label{eq:13}
  \rho_i^b = \frac{k_i}{\av{k}N}, \qquad 
  \rho_i^w = \frac{s_i}{\av{s}N},
\end{equation}
respectively in binary and weighted networks.

Other characteristic properties of the random walk, relevant to the
properties of searching in networks, are the \textit{mean
  first-passage time} (MFPT) $\tau_i$ and the \textit{coverage} $C(t)$
\cite{WeissRandomWalk,hughes,lovasz}. The MFPT of a node $i$ is
defined as the average time taken by the random walker to arrive for
the first time at $i$, starting from a random initial position in the
network. This definition gives the number of messages that have to be
exchanged, on average, in order to find vertex $i$. The coverage
$C(t)$, on the other hand, is defined as the number of different
vertices that have been visited by the walker at time $t$, averaged
for different random walks starting from different sources. The
coverage can thus be interpreted as the searching efficiency of the
network, measuring the number of different individuals that can be
reached from an arbitrary origin in a given number of time steps.

At a mean-field level, these quantities are computed as follows: let us
define $P_f(i; t)$ as the probability for the walker to arrive for the
first time at vertex $i$ in $t$ time steps. Since in the steady state
$i$ is reached in a jump with probability $\rho_i$, we have $P_f(i;t)
= [1-\rho_i]^{t-1} \rho_i$. The MFPT to vertex $i$ can thus be
estimated as the average $\tau_i = \sum_t t P_f(i;t)$, leading to
\begin{equation}
  \label{eq:17}
  \tau_i = \sum_{t=1}^\infty t [1-\rho_i]^{t-1} \equiv \frac{1}{\rho_i}.
\end{equation}
On the other hand, we can define the {\it random walk reachability} of vertex
$i$, $P_r(i;t)$, as the probability that vertex $i$ is visited by a
random walk starting at an arbitrary origin, at any time less than or
equal to $t$. The reachability takes the form 
\begin{equation}
  \label{eq:8}
  P_r(i;t)=1-[1-\rho_i]^t \simeq 1-\exp(-t \rho_i),
\end{equation}
where the last expression is valid in the limit of sufficiently small
$\rho_i$. The coverage of a random walk at time $t$ will thus be given by
the sum of these probabilities, i.e.
\begin{equation}
  \label{eq:18}
  \frac{C(t)}{N} = \frac{1}{N} \sum_i  P_r(i;t)\equiv 1-
  \frac{1}{N} \sum_i  \exp\left(-t \rho_i\right). 
\end{equation}
For sufficiently small $\rho_i t$, the
exponential in Eq.~\eqref{eq:18} can be expanded to yield $C(t)\sim
t$, a linear coverage implying that at the initial stages of the walk,
a different vertex is visited at each time step, independently of the
network properties \cite{stauffer_annealed2005,almaas03:_scaling}.

It is  now important to note that the random walk process has
been defined here in a way such that the walker performs a move and
changes node at each time step, potentially exploring a new node:
except in the pathological case of a random walk starting on an
isolated node, the walker has always a way to move out of the node it
occupies. In the context of temporal networks, on the other hand, the
walker might arrive at a node $i$ that at the successive time step
becomes isolated, and therefore has to remain trapped on that node
until a new link involving $i$ occurs. In order to compare in a
meaningful way random walk processes on static and dynamical networks,
and on different dynamical networks, we consider in each dynamical
network the average probability $\overline{p}$ that a node has at
least one link. The walker is then expected to move on average once
every $\frac{1}{\overline{p}}$ time steps, so that we will consider
the properties of the random walk process on dynamical networks as a
function of the rescaled time $\overline{p} t$.

\section{Empirical dynamical networks}
\label{sec:real-dynam-netw}

\subsection{Basics on temporal networks}

Dynamical or temporal networks \cite{Holme:2011fk} are properly
represented in terms of a \emph{contact sequence}, representing the
contacts (edges) as a function of time: a set of triplets $(i,j,t)$
where $i$ and $j$ are interacting at time $t$, with $t = \{1, \ldots,
T\}$, where $T$ is the total duration of the contact sequence. The
contact sequence can thus be expressed in terms of a
\emph{characteristic function} (or temporal adjacency matrix
\cite{Newman2010}) $\chi(i,j,t)$, taking the value $1$ when actors $i$
and $j$ are connected at time $t$, and zero otherwise.

Coarse-grained information about the structure of dynamical networks
can be obtained by projecting them onto aggregated static networks,
either binary or weighted. The binary projected network informs of the
total number of contacts of any given actor, while its weighted
version carries additional information on the total time spent in
interactions by each actor
\cite{Onnela:2007,Holme:2011fk,Isella:2011,Stehle:2011}. The
aggregated binary network is defined by an adjacency matrix of the
form
\begin{equation}
  \label{eq:9}
  a_{ij} = \Theta\left( \sum_{t} \chi(i,j,t) \right),
\end{equation}
where $\Theta(x)$ is the Heaviside theta function defined by
$\Theta(x) =1$ if $x>0$ and $\Theta(x)=0$ if $x\leq0$. In this
representation, the degree of vertex $i$, $k_i = \sum_j a_{ij}$,
represents the number of different agents with whom agent $i$ has
interacted. The associated weighted network, on the other hand, has
weights of the form
\begin{equation}
  \label{eq:10}
  \omega_{ij} = \dfrac{1}{T}\sum_{t} \chi(i,j,t).
\end{equation}
Here, $\omega_{ij}$ represents the number of interactions between
agents $i$ and $j$, normalized by its maximum possible value,
i. e. the total duration of the contact sequence $T$.  The strength of
vertex $i$, $s_i = \sum_j \omega_{ij}$, represents the average number
of interactions of agent $i$ at each time step.

While static projections represent a first step in the understanding
of the properties of dynamical networks, they coarse-grain 
a great
deal of information from the empirical time series, a fact that can be
particularly relevant when considering dynamical processes running on
top of dynamical networks \cite{Isella:2011}.  At a basic topological
level, projected networks disregard the fact that dynamics on temporal
networks are in general restricted to follow \emph{time respecting
  paths}
\cite{PhysRevE.71.046119,Kostakos:2009,Isella:2011,Nicosia:2011,Bajardi:2011,Holme:2011fk},
meaning that if a contact between vertices $i$ and $j$ took place at
times ${\cal T}_{ij} \equiv
\{t_{ij}^{(1)},t_{ij}^{(2)},\cdots,t_{ij}^{(n)} \}$, it cannot be used
in the course of a dynamical processes at any time $t \not\in {\cal
  T}_{ij}$. Therefore, not all the network is  available for
propagating a dynamics that starts at any given node, but only those
nodes belonging to its set of influence \cite{PhysRevE.71.046119},
defined as the set of nodes that can be reached from a given one,
following time respecting paths.  Moreover, an important role can also
be played by the bursty nature of dynamical and social processes,
where the appearance and disappearance of links do not follow a
Poisson processes, but show instead long tails in the distribution of
link presence and absence durations, as well as long range
correlations in the times of successive link occurrences
\cite{barabasi2005origin,10.1371/journal.pone.0011596,Gautreau:2009,Bajardi:2011}.

\subsection{Empirical contact sequences}
\label{sec:empirical-data}

The temporal networks used in the present study describe the sequences
of face-to-face contact between individuals recorded by the
SocioPatterns collaboration
\cite{Sociopatternsweb,10.1371/journal.pone.0011596}: in the
deployments of the SocioPatterns infrastructure, each individual wears
a badge equipped with an active radio-frequency identification (RFID)
device. These devices engage in bidirectional radio-communication at
very low power when they are close enough, and relay the information
about the proximity of other devices to RFID readers installed in the
environment. The devices properties are tuned so that face-to-face
proximity (1-2 meters) of individuals wearing the tags on their chests
can be assessed with a temporal resolution of $20$ seconds ($\Delta
t_0 = 20$ seconds represents thus the elementary time interval that
can be considered).

We consider here datasets describing the face-to-face proximity of
individuals gathered in several different social contexts: the
European Semantic Web Conference (``eswc''), the Hypertext conference
(``ht''), the 25th Chaos Communication Congress (``25c3'')
\footnote{In this particular case, the proximity detection range
  extended to 4-5 meters and packet exchange between devices was not
  necessarily linked to face-to-face proximity.}, and a primary school
(``school''). A description of the corresponding contexts and various
analyses of the corresponding datasets can be found in
Refs~\cite{10.1371/journal.pone.0011596,percol,Isella:2011,Stehle:2011}.

In Table~\ref{tab:summary} we summarize the main average properties of
the datasets we are considering, that are of interest in the context
of walks on dynamical networks. In particular, we focus on:
\begin{itemize}
\item $N$: number of different individuals engaged in interactions;
\item $T$: total duration of the contact sequence, in units of the
  elementary time interval $\Delta t_0 = 20$ seconds;
\item $\av{k}= \sum_i k_i / N$: average degree of nodes in the
  projected binary network, aggregated over the whole dataset;
\item $\overline{p}= \sum_t p(t)/T$: average number of individuals
  $p(t)$ interacting at each time step;
\item $\overline{f}= \sum_t E(t)/T= \sum_{ijt}\chi(i,j,t)/2T$: mean
  frequency of the interactions, defined as the average number of
  edges $E(t)$ of the instantaneous network at time $t$;
\item $\overline{n}= \sum_t n(t)$/2T: average number of new
  conversations $n(t)$ starting at each time step;
\item $\av{\Delta t_c}$: average duration of a contact.
\item $\av{s}= \sum_i s_i /N$: average strength of nodes in the
  projected weighted network, defined as the mean number of
  interactions per agent at each time step, averaged over all agents.
\end{itemize}

\begin{table}[t]
  \begin{ruledtabular} 
    \begin{tabular}{|c||c|c|c||c|c|c|c|c|}
    Dataset & $N$ & $T$ & $\av{k}$ & $\overline{p}$ & $\overline{f}$ & $\overline{n}$ &$\Delta t_c$ & $\av{s}$ \\ \hline 
      25c3 &  569 & 7450 & 185 & 0.215 & 256 & 91 & 2.82 & 0.90\\
      eswc &  173 & 4703 & 50 & 0.059 & 7 & 2.8 & 2.41 & 0.079 \\
      ht &    113 & 5093 & 39 & 0.060 & 4 & 1.9 & 2.13 & 0.072 \\
      school& 242 & 3100 & 69 & 0.235 & 41 & 25 & 1.63 & 0.34 \\
    \end{tabular}
  \end{ruledtabular} 
  \caption{Some average properties of the datasets under consideration.} \label{tab:summary}
\end{table}

Table~\ref{tab:summary} shows the heterogeneity of the considered 
datasets, in terms of size, overall duration and contact
densities. In particular, while the dataset 25c3 shows a high density
of interactions (high $\overline{p}$, $\overline{f}$ and
$\overline{n}$), and consequently a large average degree and average
strength, the others are sparser. Moreover, as also shown in the
deployments timelines in \cite{10.1371/journal.pone.0011596}, some of
the datasets show large periods of low activity, followed by bursty
peaks with a lot of contacts in few time steps, while others present
more regular interactions between elements. In this respect, it is
worth noting that we will not consider those portions of the datasets
with very low activity, in which only few couples of elements
interact, such as the beginning or ending part of conferences or the
nocturnal periods.

\begin{figure}[t]
\begin{center}
\includegraphics*[width=0.5\textwidth]{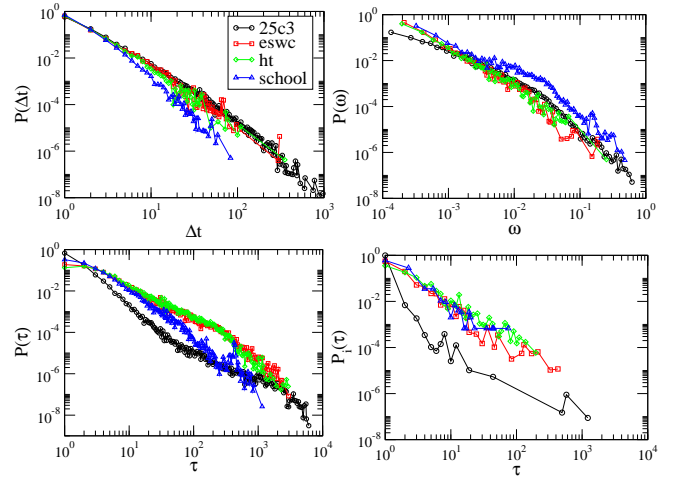}
\end{center}
\caption{(Color Online) Distributions of $P(\Delta t)$ (duration of contacts),
  $P(\omega)$ (total contact time between pairs of agents),
  $P_i(\tau)$ (gap times of a single individual $i$) and $P(\tau)$
  (global gap times).  In the case of $P_i(\tau)$, we only plot the gap times
  distribution of the agent which engages in the largest number of
  conversation, but the other agents exhibit a similar behavior.
  All distributions are heavy-tailed, indicating the bursty nature of
  face-to-face interactions, for the four empirical contact sequences
  considered.  
    } \label{fig:statistics}
\end{figure}

The heterogeneity and burstiness of the contact patterns of the
face-to-face interactions~\cite{10.1371/journal.pone.0011596} are
revealed by the study of the distribution of the duration $\Delta t$
of contacts between pairs of agents, $P(\Delta t)$, the distribution
of the total time in contact of pairs of agents (the weight
distribution $P(\omega)$), and the distribution of gap times, $\tau$, 
between two consecutive conversations involving a common
individual and two other different agents, for a single agent $i$, 
$P_i(\tau)$, or considering all the agents, $P(\tau)$.
All these distributions are heavy-tailed, typically compatible with
power-law behaviors (see Fig.~\ref{fig:statistics}), corresponding to
the burstiness of human interactions \cite{barabasi2005origin}. 

As noted above, diffusion processes such as random walks are moreover
particularly impacted by the structure of paths between nodes.  In
this respect, time respecting paths represent a crucial feature of any
temporal network, since they determine the set of possible causal
interactions between the actors of the graph.

For each (ordered) pair of nodes $(i,j)$, time-respecting paths from
$i$ to $j$ can either exist or not; moreover, the concept of shortest path on
static networks (i.e., the path with the minimum number of links
between two nodes) yields several possible generalizations in a
temporal network:
\begin{itemize}

\item the {\em fastest} path is the one that allows to go from $i$ to
  $j$, starting from the dataset initial time, in the minimum possible
  time, independently of the number of intermediate steps;

\item the {\em shortest} time-respecting path between $i$ and $j$ is
  the one that corresponds to the smallest number of intermediate
  steps, independently of the time spent between the start from $i$
  and the arrival to $j$.

\end{itemize}

For each node pair $(i,j)$, we denote by $l_{ij}^f$,
$l_{ij}^{s,temp}$, $l_{ij}^{s,stat}$ the lengths (in terms of the
number of hops) respectively of the fastest path, the shortest
time-respecting path, and the shortest path on the aggregated network,
and by $\Delta t_{ij}^f$ and $\Delta t_{ij}^s$ the duration of the
fastest and shortest time-respecting paths, where we take as initial
time the first appearance of $i$ in the dataset.  As already noted in
other works \cite{Kleinberg:2008,Isella:2011}, $l_{ij}^f$ can be much
larger than $l_{ij}^{s,stat}$. Moreover, it is clear that $l_{ij}^f
\ge l_{ij}^{s,temp} \ge l_{ij}^{s,stat}$; from the duration point of
view, on the contrary, $\Delta t_{ij}^f \le \Delta t_{ij}^s$.

We therefore define the following quantities:
\begin{itemize}
\item $l_e $: fraction of the $N(N-1)$ ordered pairs of nodes for
  which a time-respecting path exists;
\item $\langle l_s \rangle $: average length (in terms of number of
  hops along network links) of the shortest time-respecting paths;
 
\item $ \langle \Delta t_s \rangle$: average duration of the shortest
  time-respecting paths;

\item $\langle l_f \rangle $: average length of the fastest
  time-respecting paths;

\item $\langle \Delta t_f \rangle$: average duration of the fastest
  time-respecting paths;

\item $\langle l_{s,stat} \rangle$: average shortest path length in
  the binary (static) projected network;
\end{itemize}

\begin{table}[t]
  \begin{ruledtabular} 
    \begin{tabular}{|c||c|c|c|c|c|c|}
    Dataset &   $l_e$ & $\langle l_s \rangle$ & $ \langle \Delta t_s \rangle$ & $\langle l_f \rangle $ &  
$\langle \Delta t_f \rangle$ &  $\langle l_{s,stat} \rangle$ \\ \hline 
      25c3  &  0.91  & 1.67 & 1607 & 4.7  & 893 & 1.67   \\ 
      eswc  & 0.99   & 1.75 &  884 & 4.95 & 287 & 1.73   \\ 
      ht    &  0.99  & 1.67 & 1157 & 3.86 & 452 & 1.66 \\ 
      school&  1     & 1.76 & 853  & 8.27 & 349 & 1.73  \\ 
    \end{tabular}
  \end{ruledtabular} 
  \caption{(Color Online) Average properties of the shortest time-respecting paths, fastest paths and shortest 
    paths in the projected network, in the datasets considered.} \label{tab:paths}
\end{table}

The corresponding empirical values are reported in Table
\ref{tab:paths}.  It turns out that the great majority of pairs of
nodes are causally connected by at least one path in all
datasets. Hence, almost every node can potentially be influenced by
any other actor during the time evolution, i.e., the set of sources
and the set of influence of the great majority of the elements are
almost complete (of size $N$) in all of the considered datasets.

\begin{figure}[t]
\begin{center}
\includegraphics*[width=0.42\textwidth]{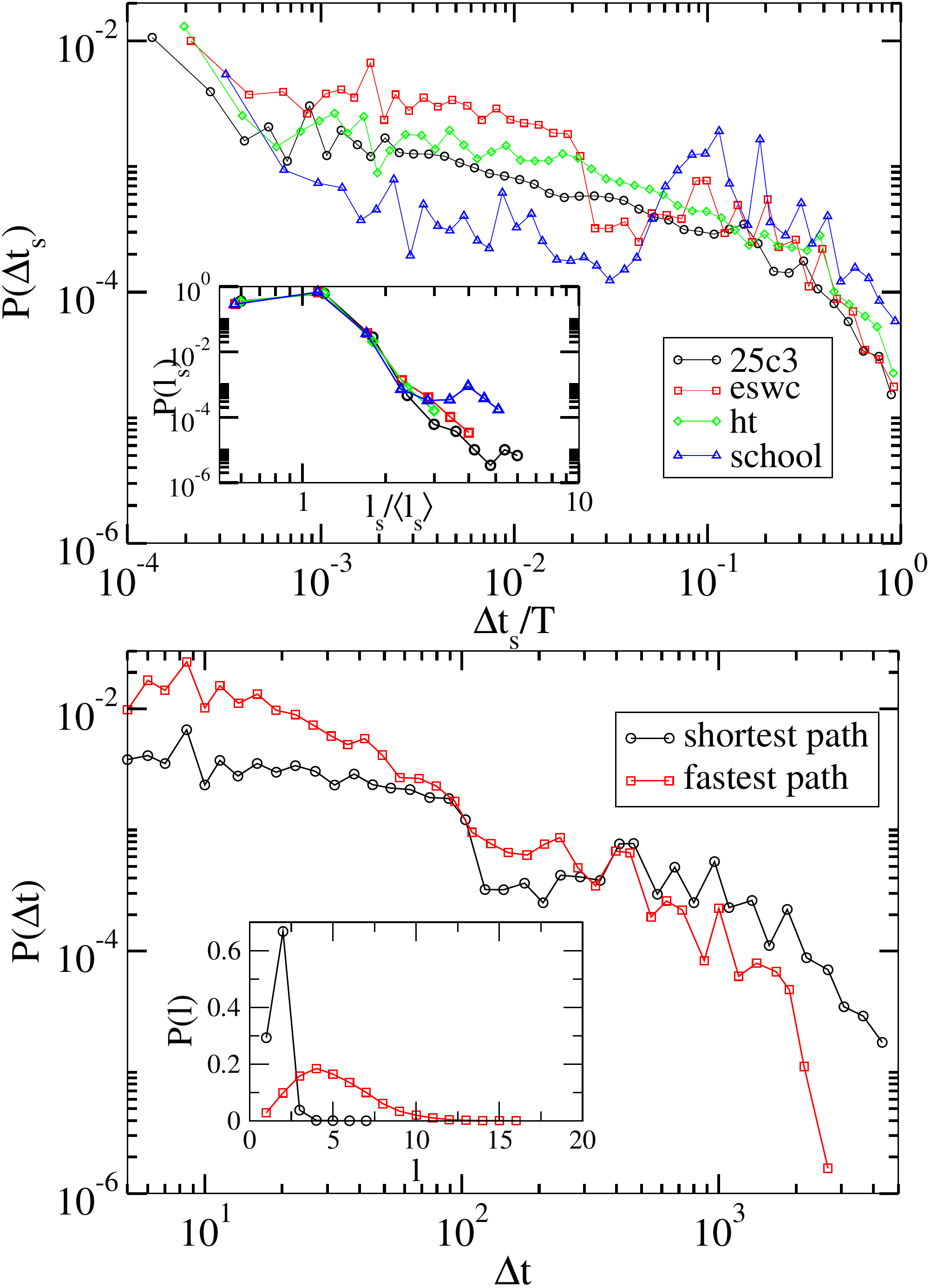}
\end{center}
\caption{(Color Online)  Top: Distribution of the temporal duration of the shortest
  time-respecting paths, normalized by its maximum value $T$. Inset:
  probability distribution $P(l_s)$ of the shortest path length measured
  over time-respecting paths, and normalized with its mean value
  $\langle l_s \rangle$. Note that the different datasets collapse.
  Bottom: Probability distribution of the duration of the shortest
  $P(\Delta t_s)$ and fastest $P(\Delta t_f)$ time-respecting paths,
  for the eswc dataset.  Inset: Probability distribution of the
  shortest $P(l_s)$ and fastest $P(l_f)$ path length for the same
  dataset.  }
  \label{fig:topodyn}
\end{figure}

In Fig. \ref{fig:topodyn} we show the distributions of the lengths,
$P(l_s)$, and durations, $P(\Delta t_s)$, of the shortest
time-respecting path for different datasets.  In the same Figure we
choose one dataset to compare the $P(l_s)$ and the $P(\Delta t_s)$
distributions with the distributions of the lengths, $P(l_f)$, and
durations, $P(\Delta t_f)$, of the fastest path.  The $P(l_s)$
distribution is short tailed and peaked on $l=2$, with a small average
value $\langle l_s \rangle $, even considering the relatively small
sizes $N$ of the datasets, and it is very similar to the projected
network one $\langle l_{s,stat} \rangle $ (see Table~\ref{tab:paths}).
The $P(l_f)$ distribution, on the contrary, shows a smooth behavior,
with an average value $\langle l_f \rangle $ several times bigger than
the shortest path one, $\langle l_s \rangle $, as expected
\cite{Kleinberg:2008,Isella:2011}.  Note that, despite the important
differences in the datasets characteristics, the $P(l_s)$
distributions (as well as $P(l_f)$, although not shown) collapse, once
rescaled.  On the other hand, the $P(\Delta t_s)$ and $P(\Delta t_f)$
distributions show the same broad-tailed behavior, but the average
duration $\langle \Delta t_s \rangle$ of the shortest paths is much
longer than the average duration $\langle \Delta t_f \rangle$ of the
fastest paths, and of the same order of magnitude than the total
duration of the contact sequence $T$.

Thus, a temporal network may be topologically
well connected and at the same time difficult to navigate or
search. Indeed spreading and searching processes need to follow paths
whose properties are determined by the temporal dynamics of the
network, and that might be either very long or very slow.

\subsection{Synthetic extensions of empirical contact sequences}
\label{sec:synth-extens-empir}

The empirical contact sequences represent the proper dynamical network
substrate upon which the properties of any dynamical process should be
studied. In many cases however, the finite duration of empirical
datasets is not sufficient to allow these processes to reach their
asymptotic state \cite{Pan:2011,Stehle:2011nx}.  This issue is
particularly important in processes that reach a steady state, such as
random walks. As discussed in Sec.~\ref{sec:overv-rand-walks}, a walker
does not move at every time step, but only with a probability
$\overline{p}$, and the effective number of movements of a walker is
of the order $T \overline{p}$. For the considered empirical sequences, 
this means that the ratio between the number of hops of
the walker and the network size, $T \overline{p} / N$, assumes values
between $3.01$ for the school case and $1.60$ for the eswc case.  Typically,
for a random walk processes such small times permit to observe
transient effects only, but not a stationary behavior. Therefore we will first
explore the asymptotic properties of random
walks in synthetically extended contact sequences, and we will consider
the corresponding finite time effects in Sec.~\ref{sec:random-walks-finite}. 
The synthetic extensions preserve at different levels the statistical properties
observed in the real data, thus providing null models of
dynamical networks. 


Inspired by previous approaches to the synthetic extension of
empirical contact sequences
\cite{PhysRevE.71.046119,Pan:2011,Stehle:2011nx,dynnetkaski2011,Holme:2011fk}, we
consider the following procedures:
\begin{itemize}
\item \textbf{SRep}: Sequence replication. The contact sequence is
  repeated periodically, defining a new extended characteristic
  function such that $\chi_e^{SRep}(i,j,t) = \chi(i,j,t \bmod
  T)$. This extension preserves all of the statistical properties of
  the empirical data (obviously, when properly rescaled to take into
  account the different durations of the extended and empirical time
  series), introducing only small corrections, at the topological
  level, on the distribution of time respecting paths and the
  associated sets of influence of each node. Indeed, a contact present
  at time $t$ will be again available to a dynamical process starting
  at time $t' > t$ after a time $t +T$.

\item \textbf{SRan}: Sequence randomization. The time ordering of the
  interactions is randomized, by constructing a new characteristic
  function such that, at each time step $t$,
  $\chi_e^{SRan}(i,j,t)=\chi(i,j,t')$ $\forall i$ and $\forall j$,
  where $t'$ is a time chosen uniformly at random from the set $\{1,
  2, \ldots, T\}$.  This form of extension yields at each time step an
  empirical instantaneous network of interactions, and preserves on
  average all the characteristics of the projected weighted network,
  but destroys the temporal correlations of successive contacts,
  leading to Poisson distributions for $P(\Delta t)$ and $P_i(\tau)$.

\item \textbf{SStat}: Statistically extended sequence. An intermediate
  level of randomization can be achieved by generating a synthetic
  contact sequence as follows: we consider the set of all
  conversations ${c(i,j,\Delta t)}$ in the sequence, defined as a
  series of consecutive contacts of length $\Delta t$ between the 
  pair of agents $i$ and $j$.  The new sequence is generated, at each
  time step $t$, by choosing $\overline{n}$ conversations
  ($\overline{n}$ being the average number of new conversations
  starting at each time step in the original sequence, see
  Table~\ref{tab:summary}), randomly selected from the set of
  conversations, and considering them as starting at time $t$ and
  ending at time $t+\Delta t$, where $\Delta t$ is the duration of the
  corresponding conversation. In this procedure we avoid choosing
  conversations between agents $i$ and $j$ which are already engaged
  in a contact started at a previous time $t' < t$. This extension
  preserves all the statistical properties of the empirical contact
  sequence, with the exception of the distribution of time gaps
  between consecutive conversations of a single individual,
  $P_i(\tau)$.
\end{itemize}

\begin{figure}[tbp]
\begin{center}
\includegraphics*[width=0.42\textwidth]{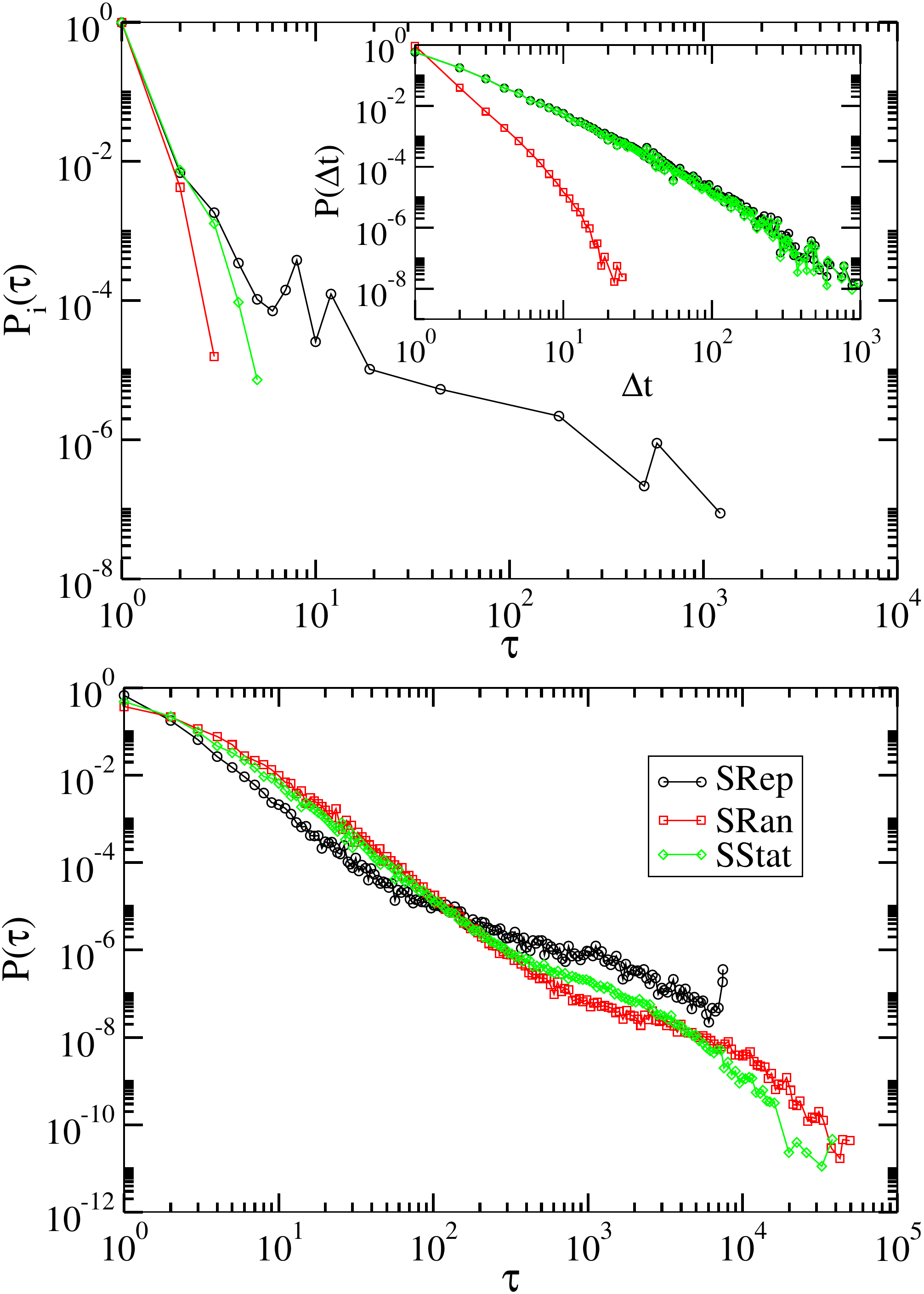}
\end{center}
\caption{(Color Online)  Top: Probability distribution $P_i(\tau)$ of a single
  individual and $P(\Delta t)$ (inset) for the extended contact
  sequences SRep, SRan and SStat, for the 25c3 dataset.  The weight
  distribution $P(w)$ of the original contact sequence is preserved
  for every extension. Bottom: Probability distribution of gap times
  $P(\tau)$ for all the agents in the SRep,  SRan and SStat extensions of the
  25c3 dataset.  }
  \label{fig:comp_extens}
\end{figure}

In Fig. \ref{fig:comp_extens} we plot the distribution of the duration
of contacts, $P(\Delta t)$, and the distribution of gap times between
two consecutive conversations realized by a single individual, $P
(\tau_i )$, for the extended contact sequences SRep, SRan and SStat.
One can check that the SRep extension preserves all the $P(w)$,
$P(\Delta t)$ and $P_i(\tau)$ distributions of the original contact
sequence, the SRan extension preserves only $P(w)$ and the SStat
extension preserves both the $P(w)$ and the $P(\Delta t)$ but not the
$P_i(\tau)$, as summarized in Table \ref{tab:extens}. Interestingly,
we note that the distribution of gap times for all agents, $P(\tau)$,
is also broadly distributed in the SRan and SStat extensions, despite the fact
that the respective individual burstiness $P_i(\tau)$ are bounded, see
Fig.~\ref{fig:comp_extens}. This fact can be easily understood by
considering that $P(\tau)$ can be written in terms of a convolution of
the individual gap distributions times the probability of starting a
conversation. In the case of SRan extension, the probability $r_i$
that an agent $i$ starts a new conversation is proportional to its
strength $s_i$, i.e. $r_i=s_i / (N \langle s \rangle) $. Therefore,
the probability that it starts a conversation $\tau$ time steps after
the last one (its gap distribution) is given by $P_i(\tau) = r_i
[1-r_i]^{\tau - 1} \simeq r_i \exp(-\tau r_i )$, for sufficiently
small $r_i$. The gap distribution for all agents $P(\tau)$ is thus
given by the convolution
\begin{equation}
  P(\tau)= \int P(s)  \dfrac{s}{N \langle s \rangle} \exp\left(-\tau
  \dfrac{s}{N \langle s \rangle} \right) ds,  
  \label{eq:14}
\end{equation}
where $P(s)$ is the strength distribution. This distribution has an
exponential form, which leads, from Eq.~\eqref{eq:14}, to a total gap
distribution $P(\tau) \sim (1+ \tau/N)^{-2}$, with a heavy
tail. Analogous arguments can be used in the case of the SStat extension.

\begin{table}[tbp]
  \begin{ruledtabular}
    \begin{tabular}{|c||  c | c  | c |}
    Extension & $P(w)$ & $P(\Delta t)$ &  $P_i(\tau)$\\ \hline 
      SRep  &   \checkmark  &  \checkmark &  \checkmark   \\ 
      SRan  &  \checkmark & \ding{55} &  \ding{55}   \\
      SStat  &   \checkmark &  \checkmark & \ding{55} \\ 
    \end{tabular}
    \end{ruledtabular} 
  \caption{Comparison of the properties of the original contact sequence preserved in the synthetic extensions.}
  \label{tab:extens}
\end{table}

\section{Random walks on extended contact sequences}
\label{sec:rand-walks-empir}

Let us consider a random walk on the sequence of instantaneous networks at
discrete time steps, which is equivalent to a message passing strategy
in which the message is passed to a randomly chosen neighbor. The
walker present at node $i$ at time $t$ hops to one of its neighbors,
randomly chosen from the set of vertices
\begin{equation}
  \label{eq:11}
  \mathcal{V}_i(t) = \left\{ j \; | \; \chi(i,j,t)=1 \right\},
\end{equation}
of which there is a number 
\begin{equation}
  \label{eq:12}
  k_i(t) = \sum_j  \chi(i,j,t),
\end{equation}
If the node $i$ is isolated at time $t$, i.e. $\mathcal{V}_i(t) =
\varnothing$, the walker remains at node $i$. 
In any case, time is increased $t\to t+1$.

Analytical considerations analogous to those in
Sec.~\ref{sec:overv-rand-walks} for the case of contact sequences are
hampered by the presence of time correlations between contacts. In
fact, as we have seen, the contacts between a given pair of agents are
neither fixed nor completely random, but instead show long range
temporal correlations. An exception is represented by the randomized
SRan extension, in which successive contacts are by construction
uncorrelated. Considering that the random walker is in vertex $i$ at
time $t$, at a subsequent time step it will be able to jump to a
vertex $j$ whenever a connection between $i$ and $j$ is created, and a
connection between $i$ and $j$ will be chosen with probability
proportional to the number of connections between $i$ and $j$ in the
original contact sequence, i.e. proportional to $\omega_{ij}$. That
is, a random walk on the extended SRan sequence behaves essentially as
in the corresponding weighted projected network, and therefore the
equations obtained in Sec.~\ref{sec:overv-rand-walks}, namely
\begin{equation}
  \label{eq:3}
  \tau_i = \frac{\av{s} N}{s_i},
\end{equation}
and
\begin{equation}
  \label{eq:4}
  \frac{C(t)}{N} = 1- \frac{1}{N} \sum_i  \exp\left(-t
    \frac{s_i}{\av{s} N}\right)
\end{equation}
apply. In this last expression for the coverage we can approximate the
sum by an integral, i.e.
\begin{equation}
  \label{eq:5}
  \frac{C(t)}{N} = 1- \int ds P(s)  \exp\left(-t
    \frac{s}{\av{s} N}\right),
\end{equation}
being $P(s)$ the distribution of strengths.  Giving that $P(s)$ has an
exponential behavior, we can obtain from the last expression
\begin{equation}
  \label{eq:6}
   \frac{C(t)}{N} \simeq 1- \left(1+\frac{t}{N}\right)^{-1}.
\end{equation}

\section{Numerical simulations}
\label{sec:numerical-simulations}

In this Section we present numerical results from the simulation of
random walks on the extended contact sequences described above.
Measuring the coverage $C(t)$ we set the duration of these sequences
to $50$ times the duration of the original contact sequence $T$, while
to evaluate the MFPT between two nodes $i$ and $j$, $\tau_{ij}$, we
let the RW explore the network up to a maximum time $t_{max}=10^8$.
Each result we report is averaged over at least $10^3$ independent
runs.

\begin{figure}[tbp]
\begin{center}
\includegraphics*[width=0.5\textwidth]{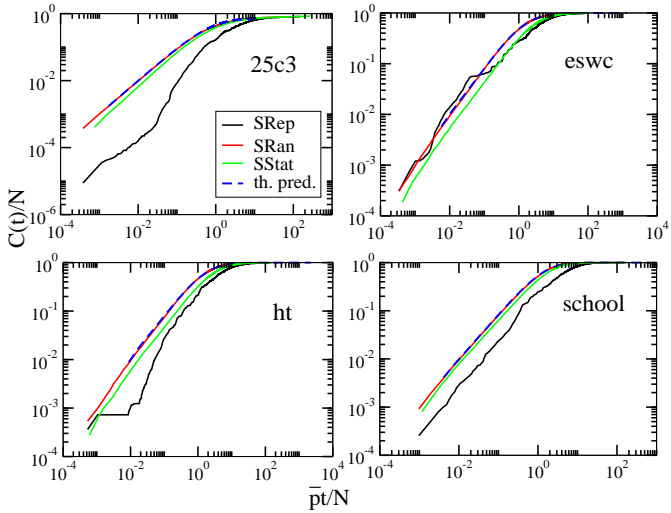}
\end{center}
\caption{(Color Online)  Normalized coverage $C(t)/N$ as a function of the rescaled
  time $\overline{p} t/N$, for the SRep, SRan and SStat extension of
  empirical data. The numerical evaluation of Eq.~(\ref{eq:4}) is
  shown as a dashed line, and each panel in the figure corresponds to
  one of the empirical datasets considered. The exploration of the
  empirical repeated data sets (SRep) is slower than the other
  cases. Moreover, the SRan is in agreement with the theoretical
  prediction, and the SStat case shows a close (but systematically
  slower) behavior. This indicates that the main slowing down factor
  in the SRep sequence is represented by the irregular distribution of
  the interactions in time, whose contribution is eliminated in the
  randomized sequences. }
  \label{fig:covextended}
\end{figure}

\subsection{Network exploration}
\label{sec:network-coverage}

The network coverage $C(t)$ describes the fraction
of nodes that the walker has discovered up to time $t$.
Figure~\ref{fig:covextended} shows the normalized coverage $C(t)/N$ as
a function of time, averaged for different walks starting from
different sources, for the dynamical networks obtained using the
 SRep, SRan and SStat prescriptions.
Time is
rescaled as $ t \to \overline{p} t$ to take into account that the walker can find 
itself on an isolated vertex, as discussed before.  While for SRep and
SRan extensions the average number of interacting nodes $\overline{p}$
is by construction the same as in the original contact sequence, for
the SStat extension we obtain numerically different values of
$\overline{p}$, which we use when rescaling time in the corresponding
simulations.

The coverage corresponding to the SRan
extension is very well fitted by a numerical simulation of
Eq.~(\ref{eq:6}), which predicts the coverage $C(t)/N$ obtained in the
correspondent projected weighted network. Moreover, when using the
rescaled time $\overline{p} t$, the SRan coverages for different
datasets collapse on top of each other for small times, with a linear
time dependence $C(t)/N \sim t/N$ for $t \ll N$ as expected in static
networks, showing a universal behavior (not shown). 

\begin{figure}[tbp]
  \begin{center}
    \includegraphics*[width=0.5\textwidth]{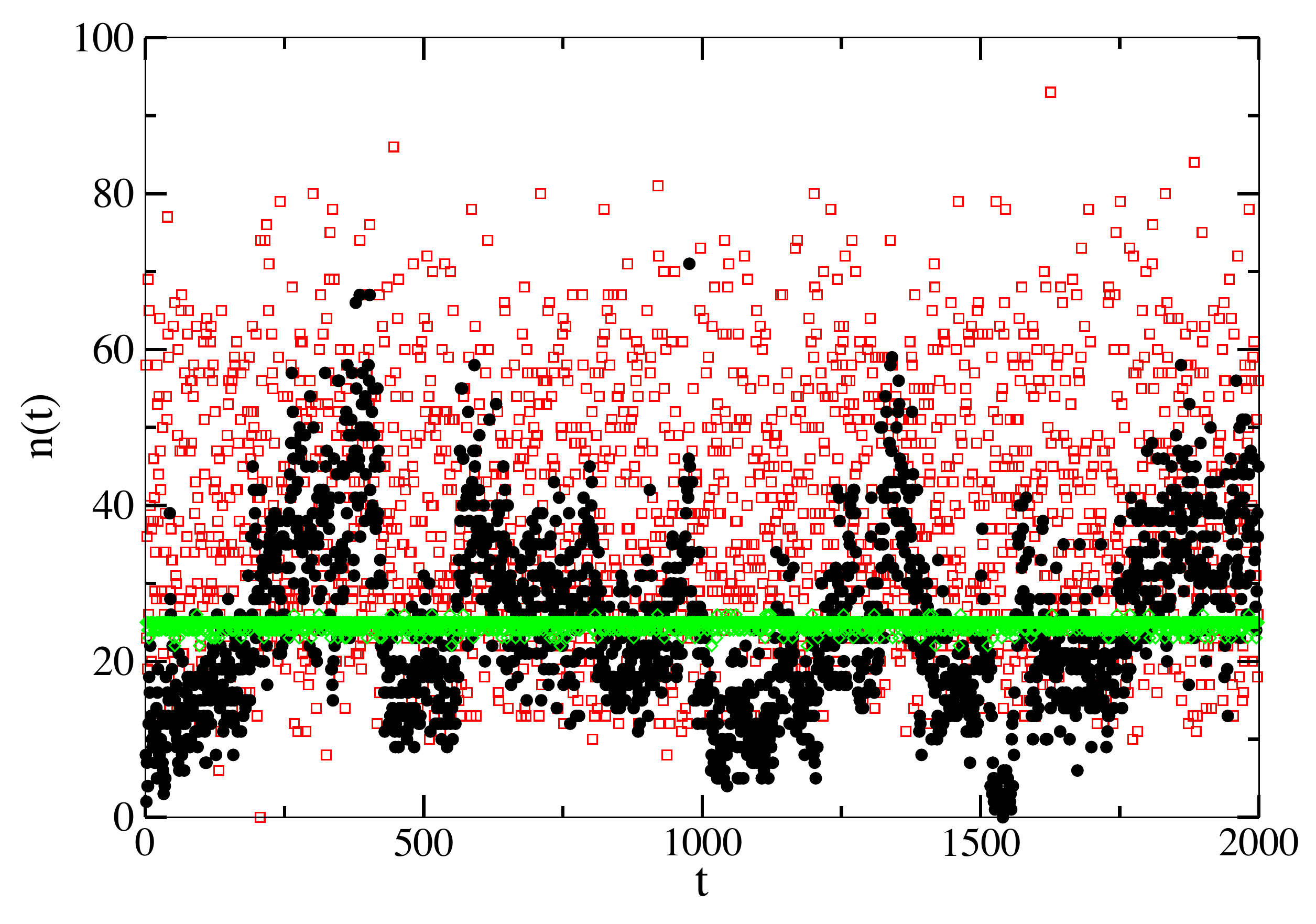}
  \end{center}
  \caption{(Color Online) Number of new conversations $n(t)$ started per unit time in the
    SRep (black, full dots), SRan (red, empty squares) and SStat (green, diamonds) extensions of the
    school dataset. 
    }
  \label{fig:numlinksschool}
\end{figure}

The coverage obtained on the SStat extension is systematically
smaller than in the SRan case, but follows a similar evolution. On the
other hand, the RW exploration obtained with the SRep prescription is
generally slower than the other two, particularly for the 25c3 and ht
datasets. As discussed before, the original contact sequence, as well
as the SRep extension, are characterized by irregular distributions of
the interactions in time, showing periods with few interacting nodes
and correspondingly a small number $n(t)$ of new started
conversations, followed by peaks with many interactions  (see
Fig.~\ref{fig:numlinksschool}). This feature slows down the RW
exploration, because the RW may remain trapped for long times on
isolated nodes. The SRan and the SStat extensions, on the contrary,
both destroy this kind of temporal structure, balancing the periods of
low and high activity: the SRan extension randomizes the time order of
the contact sequence, and the SStat extension evens the number of
interacting nodes, with $\overline{n}$ new conversations starting at
each time step.

The similarity between the random walk processes on the SRan and SStat
dynamical networks shows that the random walk coverage is not very sensitive 
to the heterogenous durations of the
conversations, as the main difference between these two cases
is that $P(\Delta t)$ is narrow for SRan and broad for SStat. In these
cases, the observed behavior is instead well accounted for by Eq. \eqref{eq:4}, 
taking into account only the weight distribution of the projected network, i.e.,
the heterogeneity between aggregated conversation durations.  Therefore, the
slower exploration properties of the SRep sequences can be
mostly attributed to the correlations between consecutive
conversations of the single individuals, as given by the individual
gap distribution $P_i(\tau)$, (see
\cite{Stehle:2011nx,Karsai:2011,dynnetkaski2011} for analogous results
in the context of epidemic spreading).

A remark is in order for the 25c3 conference. A close inspection of
Fig.~\ref{fig:covextended} shows that the RW does not reach the whole
network in any of the extensions schemes, with $C_{max} < 0.85$,
although the duration of the simulation is quite long
$\overline{p}t_{max}>10^2 N$.  The reason is that this dataset
contains a group of nodes (around $20\%$ of the total) with a very low
strength $s_i$, meaning that there are actors who are isolated for
most of the time, and whose interactions are reduced to one or two
contacts in the whole contact sequence. Given that each extension
we use preserves the $P(w)$ distribution, the discovery of these nodes
is very difficult.
The consequence is that we observe an extremely slow approach to the
asymptotic value $\lim_{t\to\infty} \frac{C(t)}{N} = 1$. Indeed, the 
mean-field calculations presented in Secs.~\ref{sec:overv-rand-walks}
and \ref{sec:synth-extens-empir} suggest a power-law decay with
$(1+\bar{p}t/N)^{-1}$ for the residual coverage $1-C(t)/N$. 
\begin{figure}[tbp]
\begin{center}
\includegraphics*[width=0.44\textwidth]{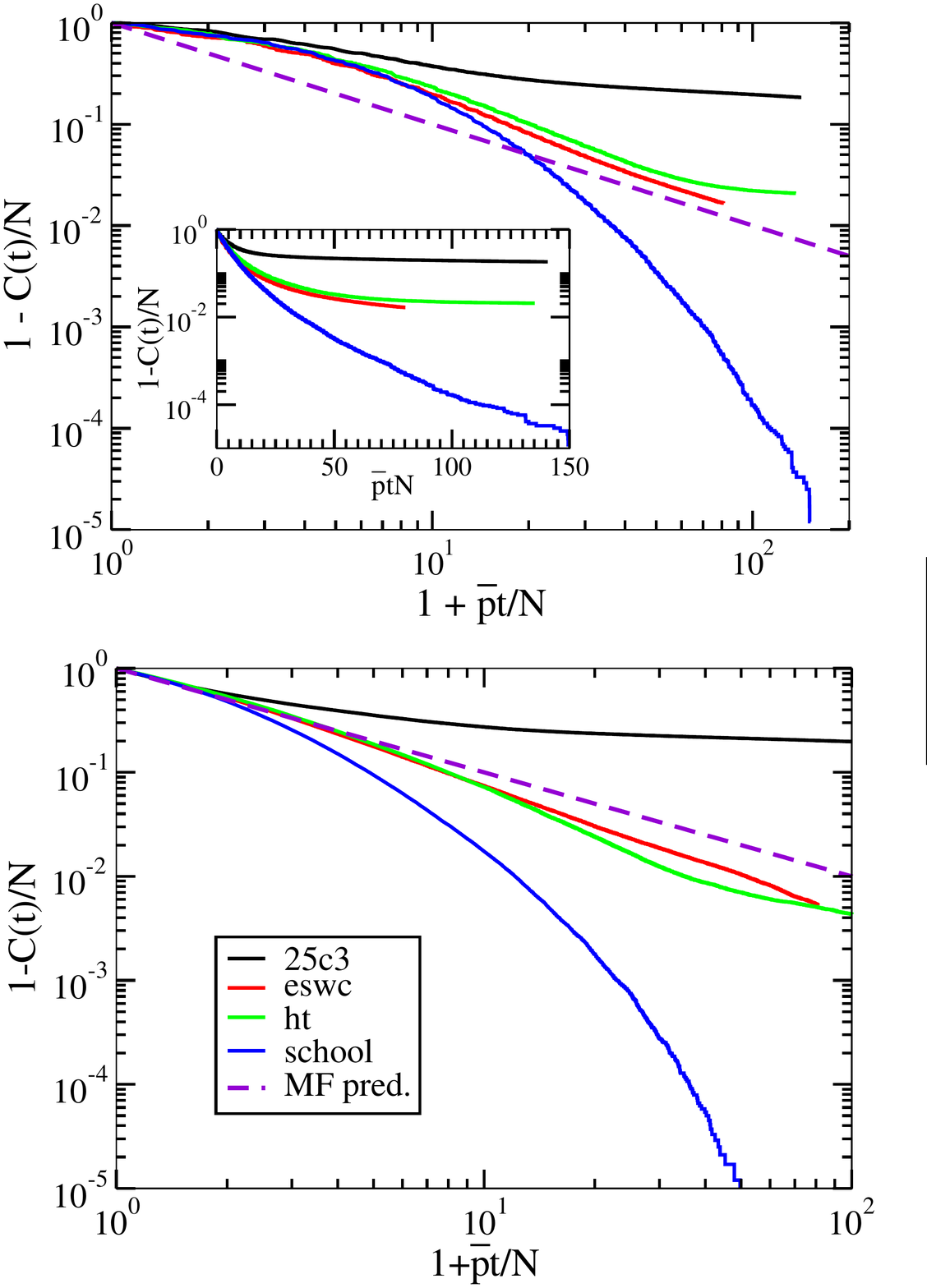}
\end{center}
\caption{(Color Online) Asymptotic residual coverage $1-C(t)/N$ as a function of $\bar{p}
  t/N$ for the SRep (top) and SRan (bottom) extended sequences, for different datasets. }
  \label{fig:residual}
\end{figure}
In Fig.~\ref{fig:residual} we plot the asymptotic coverage for large
times in the 4 datasets considered. We can see that RW on the eswc and ht dataset
conform at large times quite reasonably to the expected theoretical
prediction in Eq.~\eqref{eq:6}, both for the SRep and SRan
extensions. The 25c3 dataset shows, as discussed above, a considerable
slowing down, with a very slow decay in time. Interestingly, the school
dataset is much faster than all the rest, with a decay of the residual
coverage $1-C(t)/N$ exhibiting an approximate exponential decay. It is
noteworthy that the plots for the randomized SRan sequence do not
always obey the mean-field prediction (see lower plot in
Fig.~\ref{fig:residual}). This deviation can be attributed to the fact
that SRan extensions preserve the topological structure of the
projected weighted network, and it is known that, in some instances,
random walks on
weighted networks can deviate from the mean-field predictions
\cite{dynam_in_weigh_networ}. These deviations are particularly strong in the
case of the 25c3 dataset, where connections with a very small weight
are present.

\subsection{Mean first-passage time}
\label{sec:mean-first-passage}

\begin{figure}[tbp]
\begin{center}
\includegraphics*[width=0.49\textwidth]{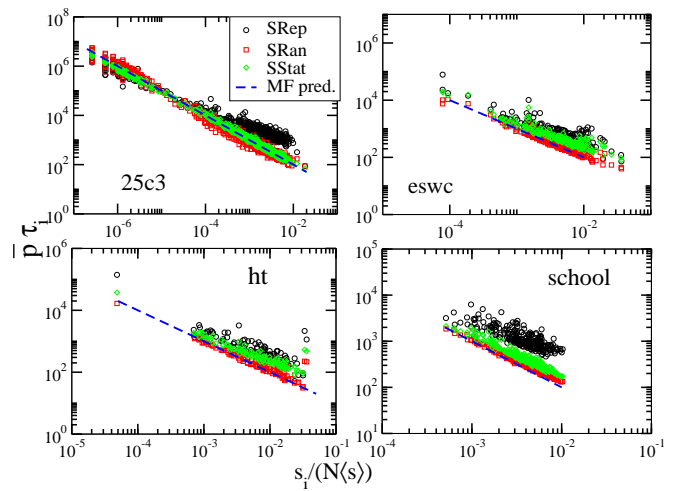}
\end{center}
\caption{(Color Online) Rescaled mean first passage time $\tau_i$,
  shown against the strength $s_i$, normalized with the total strength
  $N \langle s \rangle $, for the SRep, SRan and SStat extensions of
  empirical data. The dashed line represents the prediction of
  Eq.~\eqref{eq:3}. Each panel in the figure corresponds to one of the
  empirical datasets considered.  }
  \label{fig:mpftextended}
\end{figure}

Let us now focus on another important characteristic property of random
walk processes, namely the MFPT defined in Section
\ref{sec:overv-rand-walks}. 
Figure \ref{fig:mpftextended} shows the correlation between the MFPT $\tau_i$
of each node, measured in units of rescaled time $\overline{p} t$, and
its normalized strength $s_i/(N \langle s \rangle )$. 

\begin{figure}[tbp]
\begin{center}
\includegraphics*[width=0.49\textwidth]{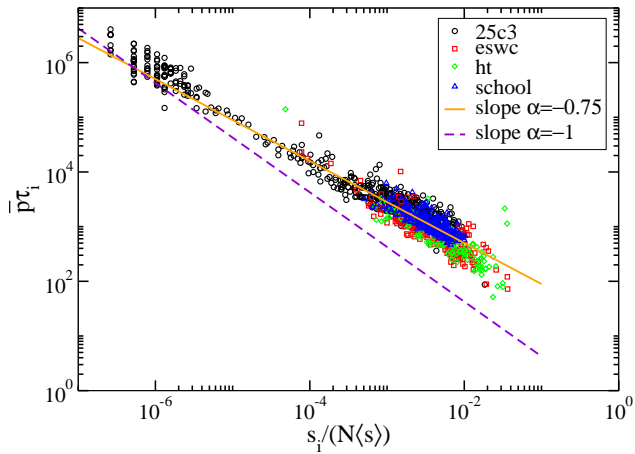}
\end{center}
  \caption{(Color Online) Mean first passage time at node $i$, in units of rescaled
    time $\overline{p} t$, vs. the strength $s_i$, normalized with the
    total strength $N \langle s \rangle $, for RW processes on the
    SRep datasets extension. All data collapse close to the continuous
    line whose slope, $\alpha \simeq 0.75$, differs from the
    theoretical one, $\alpha = 1.0$, shown as a dashed line. }
  \label{fig:mpft-rp}
\end{figure}

The random walks performed on the SRan and SStat extensions are very
well fitted by the mean field theory, i.e. Eq.~\eqref{eq:3} (predicting that $\tau_i$ 
is inversely proportional to $s_i$), for every
dataset considered; on the other hand, random walks on the extended
sequence SRep yield at the same time deviations from the mean-field
prediction and much stronger fluctuations around an average
behavior. Figure \ref{fig:mpft-rp} addresses this case in more detail,
showing that the data corresponding to RW on different datasets collapse on 
an average behavior that can be fitted by a scaling function of the form
\begin{equation}
  \label{eq:7}
  \tau_i \sim \frac{1}{\overline{p}} \times \left( \frac{s_i}{N
      \langle s \rangle 
    } \right)^{-\alpha}, 
\end{equation} 
with an exponent $\alpha \simeq 0.75$.

These results show that the MFPT, similarly to the coverage, is rather
insensitive to the distribution of the contact durations, as long as
the distribution of cumulated contact durations between individuals is
preserved (the weights of the links in the projected network). Therefore, the
deviations of the results obtained with the SRep extension of the
empirical sequences have their origin in the burstiness of the
contact patterns, as determined by the temporal correlations between
consecutive conversations. The exponent $\alpha<1$ means that the
searching process in the empirical, correlated, network is slower than
in the randomized versions, in agreement with the smaller coverage
observed in Fig.~\ref{fig:covextended}. 

The data collapse observed in Fig. \ref{fig:mpft-rp} for the SRep case leads 
to two noticeable conclusions.  First, although the various datasets studied
correspond to different contexts, with different numbers of
individuals and densities of contacts, simple rescaling procedures are
enough to compare the processes occurring on the different temporal
networks, at least for some given quantities.  Second, the MFPT at a
node is largely determined by its strength. This can indeed seem
counterintuitive as the strength is an aggregated quantity (that may
include contact events occurring at late times). However, it can be
rationalized by observing that a large strength means a large number
of contacts and therefore a large probability to be reached by the
random walker. Moreover, the fact that the strength of a node is an
aggregate view of contact events that do not occur homogeneously for
all nodes but in a bursty fashion leads to strong fluctuations around
the average behavior, which implies that nodes with the same strength can
also have rather different MFPT (Note the logarithmic scale on the
y-axis).

\section{Random walks on finite contact sequences}
\label{sec:random-walks-finite}

\begin{figure}[tbp]
\begin{center}
\includegraphics*[width=0.49\textwidth]{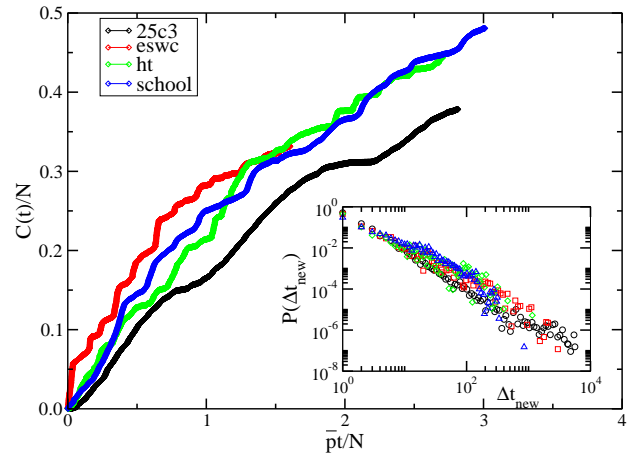}
\end{center}
  \caption{(Color Online) Normalized coverage $C(t)/N$ as function of the rescaled
    time $\overline{p} t / N$ for the different datasets.  
    The inset shows the probability distribution $P(\Delta t_{new})$ of the
    time lag $\Delta t_{new}$ between the discovery of two new vertices.
    Only the discovery of the first $5\%$ of the network is
    considered, to avoid finite size effects
    \cite{PhysRevE.78.011114}.
	  \label{fig:cov}}
\end{figure}

The case of finite sequences is interesting
from the point of view of realistic searching processes. The limited duration of a
human gathering, for example, imposes a constraint on the length of any searching strategy. 
Fig. \ref{fig:cov} shows the normalized $C(t)/N$ coverage as a
function of the rescaled time $\overline{p} t / N$. The coverage exhibits a considerable
variability in the different datasets, which do not obey the rescaling
obtained for the extended SRan and SStat sequence.  The probability distribution 
of the time lags $\Delta t_{new}$ between the discovery of two new vertices
\cite{PhysRevE.78.011114} provides further evidence
of the slowing down of diffusion in temporal networks. The inset of Fig. \ref{fig:cov} 
indeed shows broad tailed distributions $P(\Delta t_{new})$ for all the
dataset considered, differently from the exponential decay observed in
binary static networks \cite{PhysRevE.78.011114}. 

\begin{figure}[t]
\begin{center}
\includegraphics*[width=0.46\textwidth]{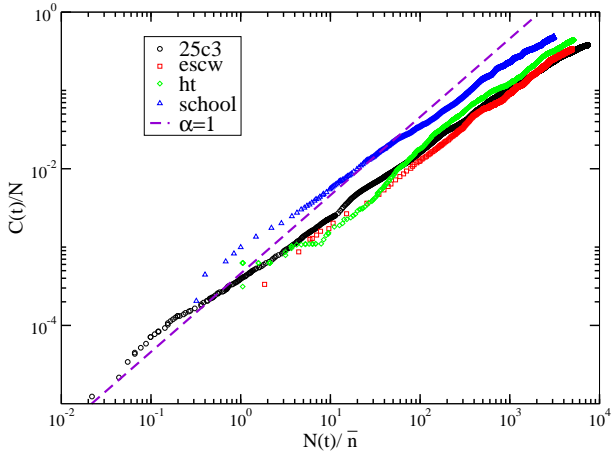}
\end{center}
\caption{(Color Online) Coverage $C(t)/N$ as a function of the number of new
  conversation realized up to time $t$, normalized for the mean number
  of new conversation per unit of time, $\overline{n}$, for different
  datasets.}
  \label{fig:linkvscov}
\end{figure}

The important differences in the rescaled coverage $C( t )/N$ between
the various datasets, shown in Fig. \ref{fig:cov}, can be attributed to
the choice of the time scale, $\overline{p}t/N$, which corresponds to
a temporal rescaling by an average quantity.  We can argue,
indeed, that the speed with which new nodes are found by the RW is
proportional to the number of new conversations $n(t)$ started at each
time step $t$, thus in the RW exploration of the temporal network the effective
time scale is given by the integrated number of new conversations up
to time $t$, $N(t) = \int^t_0 n(t') dt'$.  In Fig.~\ref{fig:linkvscov}
we display the correlation between the coverage $C(t)/N$ and the number
of new conversations realized up to time $t$, $N(t)$, normalized for
the mean number of new conversations per unit of time, $\overline{n}$.
While the relation is not strictly linear, a very strong
positive correlation appears between the two quantities.

\begin{figure}[tbp]
\begin{center}
\includegraphics*[width=0.46\textwidth]{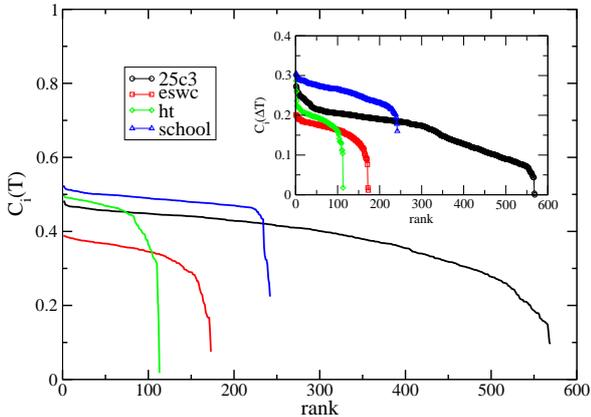}
\end{center}
  \caption{(Color Online) Rank plot of the coverage $C_i$ obtained starting from node
    $i$ in the contact sequence of duration $T$, averaged over $10^3$ runs.
    In the inset, we show a rank plot of the coverage $C_i (\Delta T)$ up to a fixed time
    $\Delta T=10^3$. }
 \label{fig:rank_cov} 
\end{figure}

The complex pattern shown by the average coverage $C(t)$
originates from the lack of self-averaging in a dynamic network. Figure \ref{fig:rank_cov}
shows the rank plot of the coverage $C_i$ obtained at the end
of a RW process starting from node $i$, and averaged over $10^3$ runs. 
Clearly, not all vertices are equivalent. 
A first explanation of the variability in $C_i$ comes from the fact that
not all nodes appear simultaneously on the network at time $0$. If
$t_{0,i}$ denotes the arrival time of node $i$ in the system, a random
walk starting from $i$ is restricted to $T^r_i =T-t_{0,i}$: nodes
arriving at later times have less possibilities to explore their set
of influence, even if this set includes all nodes. To put all nodes on
equal footing and compensate for this somehow trivial difference
between nodes, we consider the coverage of random walkers starting on
the different vertices $i$ and walking for exactly $\Delta T$ time
steps (we limit of course the study to nodes with $t_{0,i} < T- \Delta
T$). Differences in the coverage $C_i(\Delta T)$ will then depend on
the intrinsic properties of the dynamic network. For a static network
indeed, either binary or weighted, the coverage $C_i(\Delta T)$ would
be independent of $i$, as random walkers on static networks lose the
memory of their initial position in a few steps, reaching very fast
the steady state behavior Eq. \eqref{eq:13}.  As the inset of
Fig. \ref{fig:rank_cov} shows, important heterogeneities are instead
observed in the coverage of random walkers starting from different
nodes on the dynamic network, even if the random walk duration is the
same.

\begin{figure}[tbp]
\begin{center}
\includegraphics*[width=0.49\textwidth]{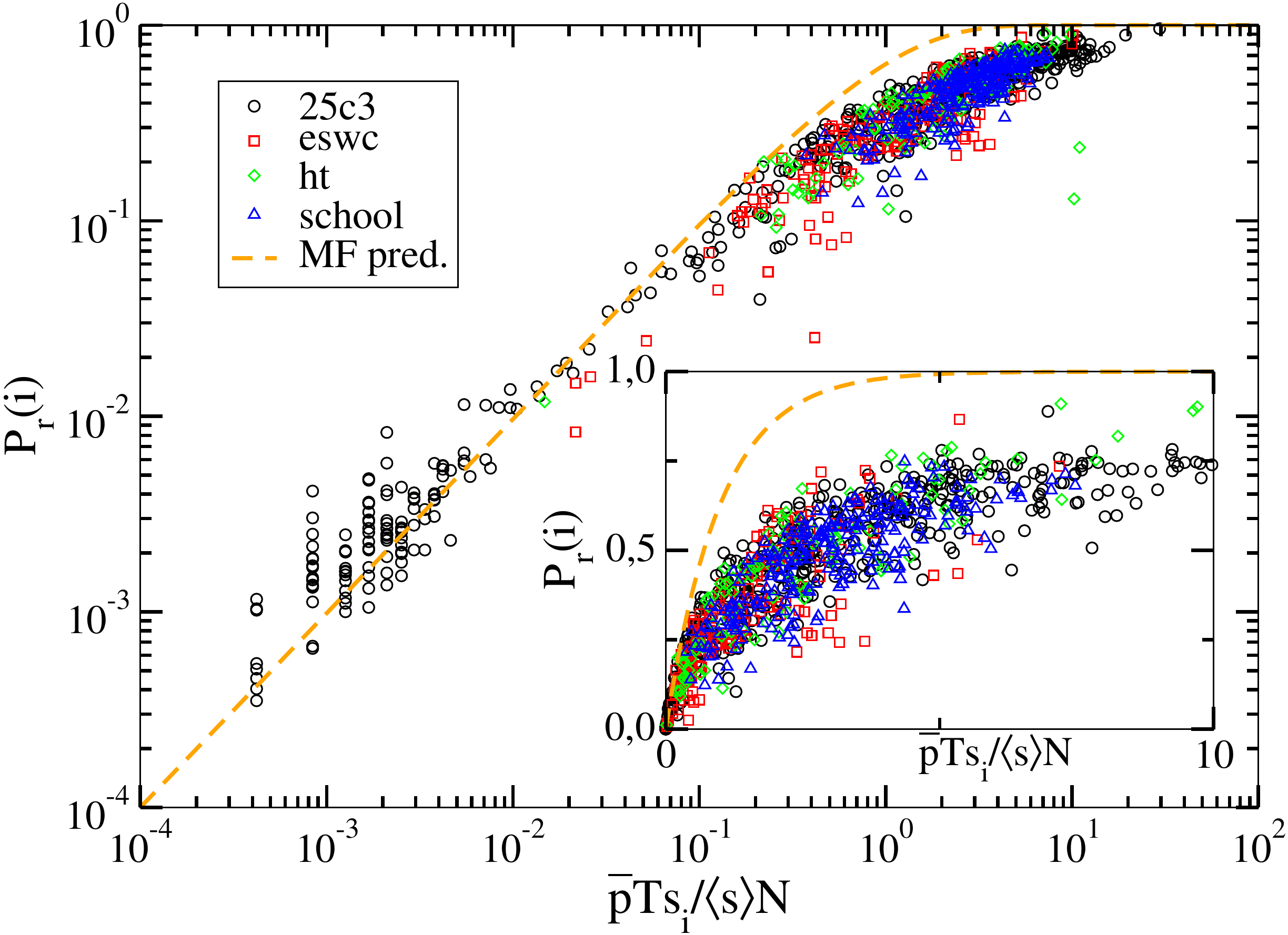}
\end{center}
  \caption{(Color Online) Correlation between the probability of node $i$ to be
    reached by the RW, $P_r(i)$, and the rescaled strength
 $\overline{p} T s_i/N \langle s \rangle $ for different datasets. The
 curves obtained by different dataset collapse, 
but they do not follow the mean-field behavior predicted by of
Equation (\ref{eq:prob_reach}) (dashed line). The inset shows the same
data on a linear scale, to emphasize the deviation from mean-field.
 \label{fig:reach-s} }
\end{figure}

Another interesting quantity is the probability that a
vertex $i$ is discovered by the random walker.
As discussed in Section
\ref{sec:overv-rand-walks}, at the mean field level the probability that a
node $i$ is visited by the RW at any time less than or equal to $t$
(the random walk reachability) takes the form $P_r (i; t) = 1 - \exp[
- t \rho (i) ]$.  Thus the probability that the node $i$ is reached by
the RW at any time in the contact sequence is
\begin{equation}
\label{eq:prob_reach}
P_r (i) = 1 - \exp \left( -\frac{\overline{p} T s_i}{N \langle s \rangle } \right),
\end{equation}
where the rescaled time $\overline{p}t $ is taken into account.  
In Fig. \ref{fig:reach-s}, we plot the probability
$P_r(i)$ of node $i$ to be reached by the RW during the contact
sequence as a function of its strength $s_i$. $P_r(i)$ exhibits a
clear increasing behavior with $s_i$,  larger strength corresponding to 
larger time in contact and therefore larger probabilities to be
reached. 
Interestingly, the simple rescaling by $\overline{p}$ and $\langle s \rangle$ leads to an
approximate data collapse for the RW processes on the various
dynamical networks, showing a very robust behavior. Similarly to the
case of the MFPT on extended sequences, the dynamical property
$P_r(i)$ can be in part ``predicted'' by an aggregate quantity such as $s_i$.
Strong deviations
from the mean-field prediction of Eq. (\ref{eq:prob_reach}) are
however observed, with a tendency of $P_r(i)$ to saturate at large
strengths to values much smaller than the ones obtained on a static
network. Thus, although the set of sources of almost every node $i$ has size
$N$, as shown in Sec. \ref{sec:empirical-data} (i.e., there exists a
time respecting path between almost every possible starting point of
the RW processes and every target node $i$), the probability for node
$i$ to be effectively reached by a RW is far from being equal to $1$.

\begin{figure}[tbp]
\begin{center}
\includegraphics*[width=0.42\textwidth]{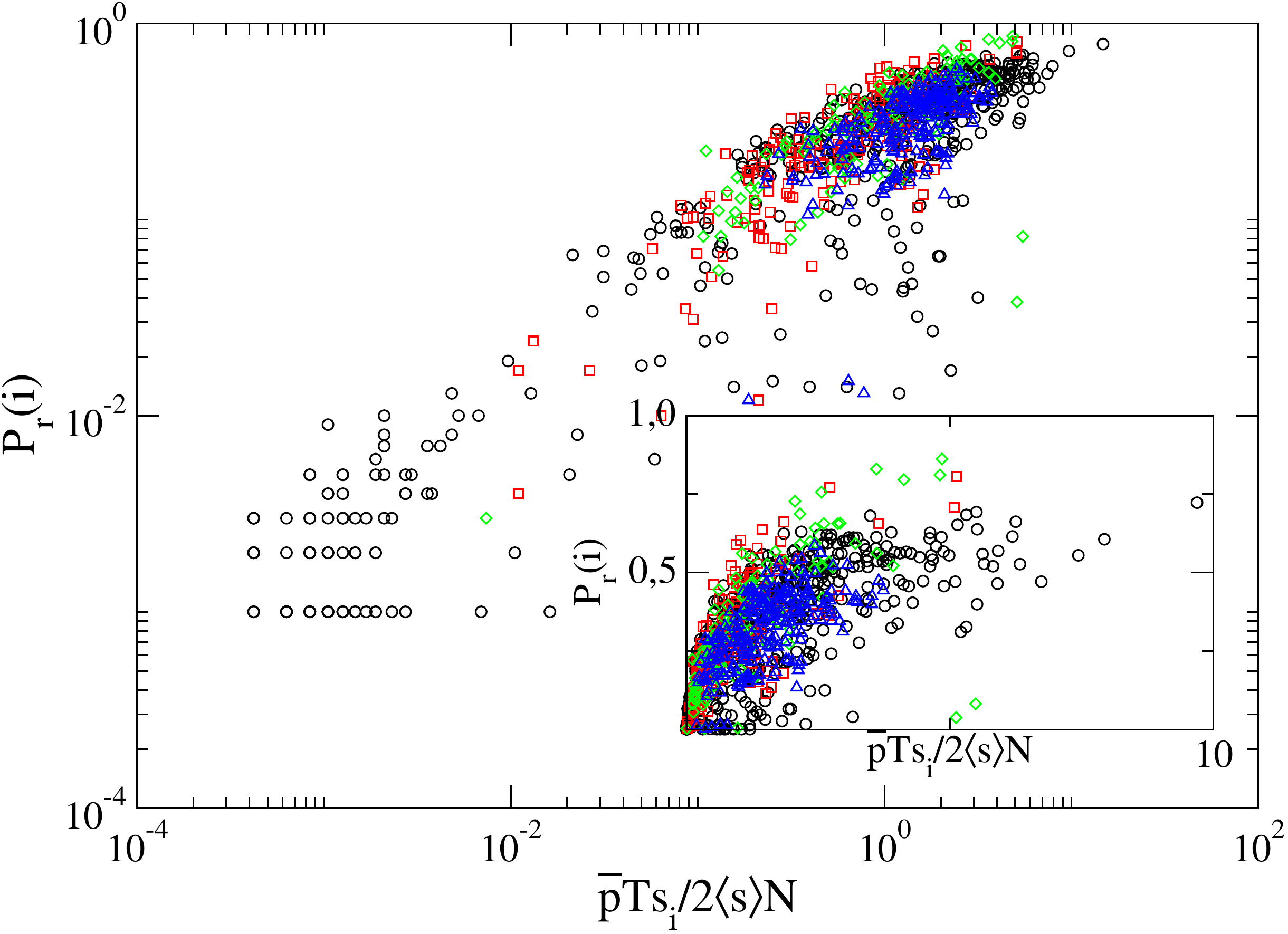}
\end{center}
  \caption{(Color Online) Correlation between the probability of node $i$ to be
    reached by a RW of length $T/2$, $P_r(i)$, and the rescaled strength
 $\overline{p} T s_i/N \langle s \rangle$ for different datasets,
 where $s_i$ is computed on the whole dataset of length $T$. The inset shows the same
data on a linear scale.
 \label{fig:reach-s-2} }
\end{figure}

Moreover, rather strong fluctuations of $P_r(i)$ at given $s_i$ are also
observed: $s_i$ is indeed an aggregate view of contacts which are typically
inhomogeneous in time, with bursty behaviors\footnote{When
  considering RW on a contact sequence of length $T$
  randomized according to the SRan procedure instead,
  Eq. (\ref{eq:prob_reach}) is well obeyed and only small fluctuations
of $P_r(i)$ are observed at a fixed $s_i$ (not shown).} . Figure \ref{fig:reach-s-2}  also shows that the
reachability computed at shorter time (here $T/2$)
displays stronger fluctuations as a function of the strength $s_i$
computed on the whole time sequence: $P_r(i)$ for shorter RW is
naturally less correlated with an aggregate view which takes into
account a more global behavior of $i$.

\section{Discussion and conclusions}
\label{sec:disc-concl}

In this paper we have investigated the behavior of random walks on
temporal networks. In particular, we have focused on real face-to-face
contact networks concerning four different datasets. These dynamical
networks exhibit heterogeneous and bursty behavior, indicated by the
long tailed distributions for the lengths and strength of
conversations, as well as for the gaps separating successive
interactions. We have underlined the importance of considering not
only the existence of time preserving paths between pairs of nodes,
but also their temporal duration: shortest paths can take much longer
than fastest paths, while fastest paths can correspond to many more
hops than shortest paths.  Interestingly, the appropriate rescaling of
these quantities identifies universal behaviors shared across the four
datasets.

Given the finite life-time of each network, we have considered as
substrate for the random walk process the replicated sequences in which
the same time series of contact patterns is indefinitely repeated. At
the same time, we have proposed two different randomization procedures
to investigate the effects of correlations in the real dataset. The
``sequence randomization" (SRan) destroys any temporal correlation by
randomizing the time ordering of the sequence. This allows to write
down exact mean-field equations for the random walker exploring these
networks, which turn out to be substantially equivalent to the ones
describing the exploration of the weighted projected network. The
``statistically extended sequence" (SStat), on the other hand, selects
random conversations from the original sequence, thus preserving the
statistical properties of the original time series, with the exception
of the distribution of time gaps between consecutive conversations.

We have performed numerical analysis both for the coverage and the
MFPT properties of the random walker. In both cases we have found that
the empirical sequences deviate systematically from the mean field
prediction, inducing a slowing down of the network exploration and of
the MFPT. Remarkably, the analysis of the randomized sequences has
allowed us to point out that this is due \textit{uniquely} to the
temporal correlations between consecutive conversations present in the
data, and \textit{not} to the heterogeneity of their lengths.  Finally, we have
addressed the role of the finite size of the empirical networks, which
turns out to prevent a full exploration of the random walker,
though differences exist across the four considered cases. In this
context, we have also shown that different starting nodes provide on
average different coverages of the networks, at odds to what happens
in static graphs. In the same way, the probability that the node $i$
is reached by the RW at any time in the contact sequence exhibits a
common behavior across the different time series, but it is not
described by the mean-field predictions for the aggregated network,
which predict a faster process.

In conclusion, the contribution of our analysis is two-fold. On the
one hand, we have proposed a general way to study dynamical processes
on temporally evolving networks, by the introduction of randomized
benchmarks and the definition of appropriate quantities that
characterize the network dynamics. On the other hand, for the
specific, yet fundamental, case of the random walk, we have obtained
detailed results that clarify the observed dynamics, and that will
represent a reference for the understanding of more complex diffusive
dynamics occurring on dynamic networks. Our investigations also open
interesting directions for future work. For instance, it would be
interesting to investigate how random walks starting from different
nodes explore first their own neighborhood \cite{baronchelli2006ring}, 
which might lead to hints about the definition of ``temporal communities'' (see
e.g. \cite{Pons:2005} for an algorithm using RW on static networks
for the detection of static communities); various measures of
nodes centrality have also been defined in temporal networks
\cite{Braha:2009,Tang:2010proc,Lerman:2010,Pan:2011,Holme:2011fk}, but their
computation is rather heavy, and RW processes might present
interesting alternatives, similarly to the case of static networks
\cite{Newman:2005}.

\section*{Acknowledgments}

We thank the SocioPatterns collaboration (www.sociopatterns.org) for providing privileged access to
dynamical network data.
M.S., R.P.-S. and A. Baronchelli acknowledge financial support from
the Spanish MEC (FEDER), under project FIS2010-21781-C02-01, and the
Junta de Andaluc\'{i}a, under project
No. P09-FQM4682.. R.P.-S. acknowledges additional support through
ICREA Academia, funded by the Generalitat de Catalunya.


%


\begin{thebibliography}{52}%
\makeatletter
\providecommand \@ifxundefined [1]{%
 \@ifx{#1\undefined}
}%
\providecommand \@ifnum [1]{%
 \ifnum #1\expandafter \@firstoftwo
 \else \expandafter \@secondoftwo
 \fi
}%
\providecommand \@ifx [1]{%
 \ifx #1\expandafter \@firstoftwo
 \else \expandafter \@secondoftwo
 \fi
}%
\providecommand \natexlab [1]{#1}%
\providecommand \enquote  [1]{``#1''}%
\providecommand \bibnamefont  [1]{#1}%
\providecommand \bibfnamefont [1]{#1}%
\providecommand \citenamefont [1]{#1}%
\providecommand \href@noop [0]{\@secondoftwo}%
\providecommand \href [0]{\begingroup \@sanitize@url \@href}%
\providecommand \@href[1]{\@@startlink{#1}\@@href}%
\providecommand \@@href[1]{\endgroup#1\@@endlink}%
\providecommand \@sanitize@url [0]{\catcode `\\12\catcode `\$12\catcode
  `\&12\catcode `\#12\catcode `\^12\catcode `\_12\catcode `\%12\relax}%
\providecommand \@@startlink[1]{}%
\providecommand \@@endlink[0]{}%
\providecommand \url  [0]{\begingroup\@sanitize@url \@url }%
\providecommand \@url [1]{\endgroup\@href {#1}{\urlprefix }}%
\providecommand \urlprefix  [0]{URL }%
\providecommand \Eprint [0]{\href }%
\providecommand \doibase [0]{http://dx.doi.org/}%
\providecommand \selectlanguage [0]{\@gobble}%
\providecommand \bibinfo  [0]{\@secondoftwo}%
\providecommand \bibfield  [0]{\@secondoftwo}%
\providecommand \translation [1]{[#1]}%
\providecommand \BibitemOpen [0]{}%
\providecommand \bibitemStop [0]{}%
\providecommand \bibitemNoStop [0]{.\EOS\space}%
\providecommand \EOS [0]{\spacefactor3000\relax}%
\providecommand \BibitemShut  [1]{\csname bibitem#1\endcsname}%
\let\auto@bib@innerbib\@empty
\bibitem [{\citenamefont {Holme}\ and\ \citenamefont
  {Saram{\"a}ki}(2011)}]{Holme:2011fk}%
  \BibitemOpen
  \bibfield  {author} {\bibinfo {author} {\bibfnamefont {P.}~\bibnamefont
  {Holme}}\ and\ \bibinfo {author} {\bibfnamefont {J.}~\bibnamefont
  {Saram{\"a}ki}},\ }\href {http://arxiv.org/abs/1108.1780v1} {\enquote
  {\bibinfo {title} {Temporal networks},}\ } (\bibinfo {year} {2011}),\ \Eprint
  {http://arxiv.org/abs/eprint arxiv:1108.1780v1} {eprint arxiv:1108.1780v1}
  \BibitemShut {NoStop}%
\bibitem [{\citenamefont {Wasserman}\ and\ \citenamefont
  {Faust}(1994)}]{wass94}%
  \BibitemOpen
  \bibfield  {author} {\bibinfo {author} {\bibfnamefont {S.}~\bibnamefont
  {Wasserman}}\ and\ \bibinfo {author} {\bibfnamefont {K.}~\bibnamefont
  {Faust}},\ }\href@noop {} {\emph {\bibinfo {title} {Social Network Analysis:
  Methods and Applications}}}\ (\bibinfo  {publisher} {Cambridge University
  Press},\ \bibinfo {address} {Cambridge},\ \bibinfo {year} {1994})\BibitemShut
  {NoStop}%
\bibitem [{\citenamefont {Newman}(2001)}]{newmancitations01}%
  \BibitemOpen
  \bibfield  {author} {\bibinfo {author} {\bibfnamefont {M.~E.~J.}\
  \bibnamefont {Newman}},\ }\href@noop {} {\bibfield  {journal} {\bibinfo
  {journal} {Proc. Natl. Acad. Sci. USA}\ }\textbf {\bibinfo {volume} {98}},\
  \bibinfo {pages} {404} (\bibinfo {year} {2001})}\BibitemShut {NoStop}%
\bibitem [{\citenamefont {Barab{\'a}si}\ and\ \citenamefont
  {Albert}(1999)}]{Barabasi:1999}%
  \BibitemOpen
  \bibfield  {author} {\bibinfo {author} {\bibfnamefont {A.-L.}\ \bibnamefont
  {Barab{\'a}si}}\ and\ \bibinfo {author} {\bibfnamefont {R.}~\bibnamefont
  {Albert}},\ }\href@noop {} {\bibfield  {journal} {\bibinfo  {journal}
  {Science}\ }\textbf {\bibinfo {volume} {286}},\ \bibinfo {pages} {509}
  (\bibinfo {year} {1999})}\BibitemShut {NoStop}%
\bibitem [{\citenamefont {Barrat}\ \emph {et~al.}(2008)\citenamefont {Barrat},
  \citenamefont {Barth\'elemy},\ and\ \citenamefont {Vespignani}}]{BBV}%
  \BibitemOpen
  \bibfield  {author} {\bibinfo {author} {\bibfnamefont {A.}~\bibnamefont
  {Barrat}}, \bibinfo {author} {\bibfnamefont {M.}~\bibnamefont
  {Barth\'elemy}}, \ and\ \bibinfo {author} {\bibfnamefont {A.}~\bibnamefont
  {Vespignani}},\ }\href@noop {} {\emph {\bibinfo {title} {Dynamical processes
  on complex networks}}}\ (\bibinfo  {publisher} {Cambridge},\ \bibinfo {year}
  {2008})\BibitemShut {NoStop}%
\bibitem [{\citenamefont {Hui}\ \emph {et~al.}(2005)\citenamefont {Hui},
  \citenamefont {Chaintreau}, \citenamefont {Scott}, \citenamefont {Gass},
  \citenamefont {Crowcroft},\ and\ \citenamefont {Diot}}]{Hui:2005}%
  \BibitemOpen
  \bibfield  {author} {\bibinfo {author} {\bibfnamefont {P.}~\bibnamefont
  {Hui}}, \bibinfo {author} {\bibfnamefont {A.}~\bibnamefont {Chaintreau}},
  \bibinfo {author} {\bibfnamefont {J.}~\bibnamefont {Scott}}, \bibinfo
  {author} {\bibfnamefont {R.}~\bibnamefont {Gass}}, \bibinfo {author}
  {\bibfnamefont {J.}~\bibnamefont {Crowcroft}}, \ and\ \bibinfo {author}
  {\bibfnamefont {C.}~\bibnamefont {Diot}},\ }in\ \href {\doibase
  http://doi.acm.org/10.1145/1080139.1080142} {\emph {\bibinfo {booktitle}
  {WDTN '05: Proceedings of the 2005 ACM SIGCOMM workshop on Delay-tolerant
  networking}}}\ (\bibinfo  {publisher} {ACM},\ \bibinfo {address} {New York,
  NY, USA},\ \bibinfo {year} {2005})\ pp.\ \bibinfo {pages}
  {244--251}\BibitemShut {NoStop}%
\bibitem [{\citenamefont {Holme}(2005)}]{PhysRevE.71.046119}%
  \BibitemOpen
  \bibfield  {author} {\bibinfo {author} {\bibfnamefont {P.}~\bibnamefont
  {Holme}},\ }\href@noop {} {\bibfield  {journal} {\bibinfo  {journal} {Phys.
  Rev. E}\ }\textbf {\bibinfo {volume} {71}},\ \bibinfo {pages} {046119}
  (\bibinfo {year} {2005})}\BibitemShut {NoStop}%
\bibitem [{\citenamefont {Onnela}\ \emph {et~al.}(2007)\citenamefont {Onnela},
  \citenamefont {Saram\"aki}, \citenamefont {Hyv\"onen}, \citenamefont
  {Szab\'o}, \citenamefont {Lazer}, \citenamefont {Kaski}, \citenamefont
  {Kert\'esz},\ and\ \citenamefont {Barab\'asi}}]{Onnela:2007}%
  \BibitemOpen
  \bibfield  {author} {\bibinfo {author} {\bibfnamefont {J.-P.}\ \bibnamefont
  {Onnela}}, \bibinfo {author} {\bibfnamefont {J.}~\bibnamefont {Saram\"aki}},
  \bibinfo {author} {\bibfnamefont {J.}~\bibnamefont {Hyv\"onen}}, \bibinfo
  {author} {\bibfnamefont {G.}~\bibnamefont {Szab\'o}}, \bibinfo {author}
  {\bibfnamefont {D.}~\bibnamefont {Lazer}}, \bibinfo {author} {\bibfnamefont
  {K.}~\bibnamefont {Kaski}}, \bibinfo {author} {\bibfnamefont
  {J.}~\bibnamefont {Kert\'esz}}, \ and\ \bibinfo {author} {\bibfnamefont
  {A.-L.}\ \bibnamefont {Barab\'asi}},\ }\href {\doibase
  10.1073/pnas.0610245104} {\bibfield  {journal} {\bibinfo  {journal}
  {Proceedings of the National Academy of Sciences}\ }\textbf {\bibinfo
  {volume} {104}},\ \bibinfo {pages} {7332} (\bibinfo {year}
  {2007})}\BibitemShut {NoStop}%
\bibitem [{\citenamefont {Gautreau}\ \emph {et~al.}(2009)\citenamefont
  {Gautreau}, \citenamefont {Barrat},\ and\ \citenamefont
  {Barth\'elemy}}]{Gautreau:2009}%
  \BibitemOpen
  \bibfield  {author} {\bibinfo {author} {\bibfnamefont {A.}~\bibnamefont
  {Gautreau}}, \bibinfo {author} {\bibfnamefont {A.}~\bibnamefont {Barrat}}, \
  and\ \bibinfo {author} {\bibfnamefont {M.}~\bibnamefont {Barth\'elemy}},\
  }\href {\doibase 10.1073/pnas.0811113106} {\bibfield  {journal} {\bibinfo
  {journal} {Proceedings of the National Academy of Sciences}\ }\textbf
  {\bibinfo {volume} {106}},\ \bibinfo {pages} {8847} (\bibinfo {year}
  {2009})}\BibitemShut {NoStop}%
\bibitem [{\citenamefont {Cattuto}\ \emph {et~al.}(2010)\citenamefont
  {Cattuto}, \citenamefont {Van~den Broeck}, \citenamefont {Barrat},
  \citenamefont {Colizza}, \citenamefont {Pinton},\ and\ \citenamefont
  {Vespignani}}]{10.1371/journal.pone.0011596}%
  \BibitemOpen
  \bibfield  {author} {\bibinfo {author} {\bibfnamefont {C.}~\bibnamefont
  {Cattuto}}, \bibinfo {author} {\bibfnamefont {W.}~\bibnamefont {Van~den
  Broeck}}, \bibinfo {author} {\bibfnamefont {A.}~\bibnamefont {Barrat}},
  \bibinfo {author} {\bibfnamefont {V.}~\bibnamefont {Colizza}}, \bibinfo
  {author} {\bibfnamefont {J.-F.}\ \bibnamefont {Pinton}}, \ and\ \bibinfo
  {author} {\bibfnamefont {A.}~\bibnamefont {Vespignani}},\ }\href@noop {}
  {\bibfield  {journal} {\bibinfo  {journal} {PLoS ONE}\ }\textbf {\bibinfo
  {volume} {5}},\ \bibinfo {pages} {e11596} (\bibinfo {year}
  {2010})}\BibitemShut {NoStop}%
\bibitem [{\citenamefont {Tang}\ \emph
  {et~al.}(2010{\natexlab{a}})\citenamefont {Tang}, \citenamefont {Scellato},
  \citenamefont {Musolesi}, \citenamefont {Mascolo},\ and\ \citenamefont
  {Latora}}]{Tang:2010}%
  \BibitemOpen
  \bibfield  {author} {\bibinfo {author} {\bibfnamefont {J.}~\bibnamefont
  {Tang}}, \bibinfo {author} {\bibfnamefont {S.}~\bibnamefont {Scellato}},
  \bibinfo {author} {\bibfnamefont {M.}~\bibnamefont {Musolesi}}, \bibinfo
  {author} {\bibfnamefont {C.}~\bibnamefont {Mascolo}}, \ and\ \bibinfo
  {author} {\bibfnamefont {V.}~\bibnamefont {Latora}},\ }\href {\doibase
  10.1103/PhysRevE.81.055101} {\bibfield  {journal} {\bibinfo  {journal} {Phys.
  Rev. E}\ }\textbf {\bibinfo {volume} {81}},\ \bibinfo {pages} {055101}
  (\bibinfo {year} {2010}{\natexlab{a}})}\BibitemShut {NoStop}%
\bibitem [{\citenamefont {Bajardi}\ \emph {et~al.}(2011)\citenamefont
  {Bajardi}, \citenamefont {Barrat}, \citenamefont {Natale}, \citenamefont
  {Savini},\ and\ \citenamefont {Colizza}}]{Bajardi:2011}%
  \BibitemOpen
  \bibfield  {author} {\bibinfo {author} {\bibfnamefont {P.}~\bibnamefont
  {Bajardi}}, \bibinfo {author} {\bibfnamefont {A.}~\bibnamefont {Barrat}},
  \bibinfo {author} {\bibfnamefont {F.}~\bibnamefont {Natale}}, \bibinfo
  {author} {\bibfnamefont {L.}~\bibnamefont {Savini}}, \ and\ \bibinfo {author}
  {\bibfnamefont {V.}~\bibnamefont {Colizza}},\ }\href {\doibase
  10.1371/journal.pone.0019869} {\bibfield  {journal} {\bibinfo  {journal}
  {PLoS ONE}\ }\textbf {\bibinfo {volume} {6}},\ \bibinfo {pages} {e19869}
  (\bibinfo {year} {2011})}\BibitemShut {NoStop}%
\bibitem [{\citenamefont {Stehl\'e}\ \emph
  {et~al.}(2011{\natexlab{a}})\citenamefont {Stehl\'e}, \citenamefont {Voirin},
  \citenamefont {Barrat}, \citenamefont {Cattuto}, \citenamefont {Colizza},
  \citenamefont {Isella}, \citenamefont {R\'egis}, \citenamefont {Pinton},
  \citenamefont {Khanafer}, \citenamefont {Van~den Broeck},\ and\ \citenamefont
  {Vanhems}}]{Stehle:2011nx}%
  \BibitemOpen
  \bibfield  {author} {\bibinfo {author} {\bibfnamefont {J.}~\bibnamefont
  {Stehl\'e}}, \bibinfo {author} {\bibfnamefont {N.}~\bibnamefont {Voirin}},
  \bibinfo {author} {\bibfnamefont {A.}~\bibnamefont {Barrat}}, \bibinfo
  {author} {\bibfnamefont {C.}~\bibnamefont {Cattuto}}, \bibinfo {author}
  {\bibfnamefont {V.}~\bibnamefont {Colizza}}, \bibinfo {author} {\bibfnamefont
  {L.}~\bibnamefont {Isella}}, \bibinfo {author} {\bibfnamefont
  {C.}~\bibnamefont {R\'egis}}, \bibinfo {author} {\bibfnamefont {J.-F.}\
  \bibnamefont {Pinton}}, \bibinfo {author} {\bibfnamefont {N.}~\bibnamefont
  {Khanafer}}, \bibinfo {author} {\bibfnamefont {W.}~\bibnamefont {Van~den
  Broeck}}, \ and\ \bibinfo {author} {\bibfnamefont {P.}~\bibnamefont
  {Vanhems}},\ }\href {http://www.biomedcentral.com/1741-7015/9/87} {\bibfield
  {journal} {\bibinfo  {journal} {BMC Medicine}\ }\textbf {\bibinfo {volume}
  {9}} (\bibinfo {year} {2011}{\natexlab{a}})}\BibitemShut {NoStop}%
\bibitem [{\citenamefont {Miritello}\ \emph {et~al.}(2011)\citenamefont
  {Miritello}, \citenamefont {Moro},\ and\ \citenamefont
  {Lara}}]{Miritello:2011}%
  \BibitemOpen
  \bibfield  {author} {\bibinfo {author} {\bibfnamefont {G.}~\bibnamefont
  {Miritello}}, \bibinfo {author} {\bibfnamefont {E.}~\bibnamefont {Moro}}, \
  and\ \bibinfo {author} {\bibfnamefont {R.}~\bibnamefont {Lara}},\ }\href
  {\doibase 10.1103/PhysRevE.83.045102} {\bibfield  {journal} {\bibinfo
  {journal} {Phys. Rev. E}\ }\textbf {\bibinfo {volume} {83}},\ \bibinfo
  {pages} {045102} (\bibinfo {year} {2011})}\BibitemShut {NoStop}%
\bibitem [{\citenamefont {Karsai}\ \emph {et~al.}(2011)\citenamefont {Karsai},
  \citenamefont {Kivel\"a}, \citenamefont {Pan}, \citenamefont {Kaski},
  \citenamefont {Kert\'esz}, \citenamefont {Barab\'asi},\ and\ \citenamefont
  {Saram\"aki}}]{Karsai:2011}%
  \BibitemOpen
  \bibfield  {author} {\bibinfo {author} {\bibfnamefont {M.}~\bibnamefont
  {Karsai}}, \bibinfo {author} {\bibfnamefont {M.}~\bibnamefont {Kivel\"a}},
  \bibinfo {author} {\bibfnamefont {R.~K.}\ \bibnamefont {Pan}}, \bibinfo
  {author} {\bibfnamefont {K.}~\bibnamefont {Kaski}}, \bibinfo {author}
  {\bibfnamefont {J.}~\bibnamefont {Kert\'esz}}, \bibinfo {author}
  {\bibfnamefont {A.-L.}\ \bibnamefont {Barab\'asi}}, \ and\ \bibinfo {author}
  {\bibfnamefont {J.}~\bibnamefont {Saram\"aki}},\ }\href {\doibase
  10.1103/PhysRevE.83.025102} {\bibfield  {journal} {\bibinfo  {journal} {Phys.
  Rev. E}\ }\textbf {\bibinfo {volume} {83}},\ \bibinfo {pages} {025102}
  (\bibinfo {year} {2011})}\BibitemShut {NoStop}%
\bibitem [{\citenamefont {Scherrer}\ \emph {et~al.}(2008)\citenamefont
  {Scherrer}, \citenamefont {Borgnat}, \citenamefont {Fleury}, \citenamefont
  {Guillaume},\ and\ \citenamefont {Robardet}}]{Scherrer:2008}%
  \BibitemOpen
  \bibfield  {author} {\bibinfo {author} {\bibfnamefont {A.}~\bibnamefont
  {Scherrer}}, \bibinfo {author} {\bibfnamefont {P.}~\bibnamefont {Borgnat}},
  \bibinfo {author} {\bibfnamefont {E.}~\bibnamefont {Fleury}}, \bibinfo
  {author} {\bibfnamefont {J.-L.}\ \bibnamefont {Guillaume}}, \ and\ \bibinfo
  {author} {\bibfnamefont {C.}~\bibnamefont {Robardet}},\ }\href@noop {}
  {\bibfield  {journal} {\bibinfo  {journal} {Comp. Net.}\ }\textbf {\bibinfo
  {volume} {52}},\ \bibinfo {pages} {2842} (\bibinfo {year}
  {2008})}\BibitemShut {NoStop}%
\bibitem [{\citenamefont {Hill}\ and\ \citenamefont {Braha}(2010)}]{Hill:2009}%
  \BibitemOpen
  \bibfield  {author} {\bibinfo {author} {\bibfnamefont {S.}~\bibnamefont
  {Hill}}\ and\ \bibinfo {author} {\bibfnamefont {D.}~\bibnamefont {Braha}},\
  }\href@noop {} {\bibfield  {journal} {\bibinfo  {journal} {Phys. Rev. E}\
  }\textbf {\bibinfo {volume} {82}},\ \bibinfo {pages} {046105} (\bibinfo
  {year} {2010})}\BibitemShut {NoStop}%
\bibitem [{\citenamefont {Stehl\'e}\ \emph {et~al.}(2010)\citenamefont
  {Stehl\'e}, \citenamefont {Barrat},\ and\ \citenamefont
  {Bianconi}}]{PhysRevE.81.035101}%
  \BibitemOpen
  \bibfield  {author} {\bibinfo {author} {\bibfnamefont {J.}~\bibnamefont
  {Stehl\'e}}, \bibinfo {author} {\bibfnamefont {A.}~\bibnamefont {Barrat}}, \
  and\ \bibinfo {author} {\bibfnamefont {G.}~\bibnamefont {Bianconi}},\
  }\href@noop {} {\bibfield  {journal} {\bibinfo  {journal} {Phys. Rev. E}\
  }\textbf {\bibinfo {volume} {81}},\ \bibinfo {pages} {035101} (\bibinfo
  {year} {2010})}\BibitemShut {NoStop}%
\bibitem [{\citenamefont {Zhao}\ \emph {et~al.}(2011)\citenamefont {Zhao},
  \citenamefont {Stehl\'e}, \citenamefont {Bianconi},\ and\ \citenamefont
  {Barrat}}]{PhysRevE.83.056109}%
  \BibitemOpen
  \bibfield  {author} {\bibinfo {author} {\bibfnamefont {K.}~\bibnamefont
  {Zhao}}, \bibinfo {author} {\bibfnamefont {J.}~\bibnamefont {Stehl\'e}},
  \bibinfo {author} {\bibfnamefont {G.}~\bibnamefont {Bianconi}}, \ and\
  \bibinfo {author} {\bibfnamefont {A.}~\bibnamefont {Barrat}},\ }\href@noop {}
  {\bibfield  {journal} {\bibinfo  {journal} {Phys. Rev. E}\ }\textbf {\bibinfo
  {volume} {83}},\ \bibinfo {pages} {056109} (\bibinfo {year}
  {2011})}\BibitemShut {NoStop}%
\bibitem [{\citenamefont {Rocha}\ \emph {et~al.}(2011)\citenamefont {Rocha},
  \citenamefont {Liljeros},\ and\ \citenamefont {Holme}}]{Rocha:2010}%
  \BibitemOpen
  \bibfield  {author} {\bibinfo {author} {\bibfnamefont {L.~E.~C.}\
  \bibnamefont {Rocha}}, \bibinfo {author} {\bibfnamefont {F.}~\bibnamefont
  {Liljeros}}, \ and\ \bibinfo {author} {\bibfnamefont {P.}~\bibnamefont
  {Holme}},\ }\href {\doibase 10.1371/journal.pcbi.1001109} {\bibfield
  {journal} {\bibinfo  {journal} {PLoS Comput Biol}\ }\textbf {\bibinfo
  {volume} {7}},\ \bibinfo {pages} {e1001109} (\bibinfo {year}
  {2011})}\BibitemShut {NoStop}%
\bibitem [{\citenamefont {Isella}\ \emph {et~al.}(2011)\citenamefont {Isella},
  \citenamefont {Stehl\'e}, \citenamefont {Barrat}, \citenamefont {Cattuto},
  \citenamefont {Pinton},\ and\ \citenamefont {den Broeck}}]{Isella:2011}%
  \BibitemOpen
  \bibfield  {author} {\bibinfo {author} {\bibfnamefont {L.}~\bibnamefont
  {Isella}}, \bibinfo {author} {\bibfnamefont {J.}~\bibnamefont {Stehl\'e}},
  \bibinfo {author} {\bibfnamefont {A.}~\bibnamefont {Barrat}}, \bibinfo
  {author} {\bibfnamefont {C.}~\bibnamefont {Cattuto}}, \bibinfo {author}
  {\bibfnamefont {J.-F.}\ \bibnamefont {Pinton}}, \ and\ \bibinfo {author}
  {\bibfnamefont {W.~V.}\ \bibnamefont {den Broeck}},\ }\href@noop {}
  {\bibfield  {journal} {\bibinfo  {journal} {J. Theor. Biol}\ }\textbf
  {\bibinfo {volume} {271}},\ \bibinfo {pages} {166} (\bibinfo {year}
  {2011})}\BibitemShut {NoStop}%
\bibitem [{\citenamefont {Kivela}\ \emph {et~al.}(2011)\citenamefont {Kivela},
  \citenamefont {{Kumar Pan}}, \citenamefont {Kaski}, \citenamefont {Kertesz},
  \citenamefont {Saramaki},\ and\ \citenamefont {Karsai}}]{dynnetkaski2011}%
  \BibitemOpen
  \bibfield  {author} {\bibinfo {author} {\bibfnamefont {M.}~\bibnamefont
  {Kivela}}, \bibinfo {author} {\bibfnamefont {R.}~\bibnamefont {{Kumar Pan}}},
  \bibinfo {author} {\bibfnamefont {K.}~\bibnamefont {Kaski}}, \bibinfo
  {author} {\bibfnamefont {J.}~\bibnamefont {Kertesz}}, \bibinfo {author}
  {\bibfnamefont {J.}~\bibnamefont {Saramaki}}, \ and\ \bibinfo {author}
  {\bibfnamefont {M.}~\bibnamefont {Karsai}},\ }\href@noop {} {\enquote
  {\bibinfo {title} {Multiscale analysis of spreading in a large communication
  network},}\ } (\bibinfo {year} {2011}),\ \bibinfo {note}
  {arXiv:1112.4312v1}\BibitemShut {NoStop}%
\bibitem [{\citenamefont {Fujiwara}\ \emph {et~al.}(2011)\citenamefont
  {Fujiwara}, \citenamefont {Kurths},\ and\ \citenamefont
  {D{\'\i}az-Guilera}}]{albert2011sync}%
  \BibitemOpen
  \bibfield  {author} {\bibinfo {author} {\bibfnamefont {N.}~\bibnamefont
  {Fujiwara}}, \bibinfo {author} {\bibfnamefont {J.}~\bibnamefont {Kurths}}, \
  and\ \bibinfo {author} {\bibfnamefont {A.}~\bibnamefont
  {D{\'\i}az-Guilera}},\ }\href@noop {} {\bibfield  {journal} {\bibinfo
  {journal} {Physical Review E}\ }\textbf {\bibinfo {volume} {83}},\ \bibinfo
  {pages} {025101} (\bibinfo {year} {2011})}\BibitemShut {NoStop}%
\bibitem [{\citenamefont {Parshani}\ \emph {et~al.}(2010)\citenamefont
  {Parshani}, \citenamefont {Dickison}, \citenamefont {Cohen}, \citenamefont
  {Stanley},\ and\ \citenamefont {Havlin}}]{Parshani:2010}%
  \BibitemOpen
  \bibfield  {author} {\bibinfo {author} {\bibfnamefont {R.}~\bibnamefont
  {Parshani}}, \bibinfo {author} {\bibfnamefont {M.}~\bibnamefont {Dickison}},
  \bibinfo {author} {\bibfnamefont {R.}~\bibnamefont {Cohen}}, \bibinfo
  {author} {\bibfnamefont {H.~E.}\ \bibnamefont {Stanley}}, \ and\ \bibinfo
  {author} {\bibfnamefont {S.}~\bibnamefont {Havlin}},\ }\href
  {http://stacks.iop.org/0295-5075/90/i=3/a=38004} {\bibfield  {journal}
  {\bibinfo  {journal} {EPL (Europhysics Letters)}\ }\textbf {\bibinfo {volume}
  {90}},\ \bibinfo {pages} {38004} (\bibinfo {year} {2010})}\BibitemShut
  {NoStop}%
\bibitem [{\citenamefont {Baronchelli}\ and\ \citenamefont
  {D{\'\i}az-Guilera}(2012)}]{consensus_temporal_nrets_2012}%
  \BibitemOpen
  \bibfield  {author} {\bibinfo {author} {\bibfnamefont {A.}~\bibnamefont
  {Baronchelli}}\ and\ \bibinfo {author} {\bibfnamefont {A.}~\bibnamefont
  {D{\'\i}az-Guilera}},\ }\href@noop {} {\bibfield  {journal} {\bibinfo
  {journal} {Phys. Rev. E}\ }\textbf {\bibinfo {volume} {85}},\ \bibinfo
  {pages} {016113} (\bibinfo {year} {2012})}\BibitemShut {NoStop}%
\bibitem [{\citenamefont {Weiss}(1994)}]{WeissRandomWalk}%
  \BibitemOpen
  \bibfield  {author} {\bibinfo {author} {\bibfnamefont {G.~H.}\ \bibnamefont
  {Weiss}},\ }\href@noop {} {\emph {\bibinfo {title} {{Aspects and Applications
  of the Random Walk}}}}\ (\bibinfo  {publisher} {North-Holland Publishing
  Co.},\ \bibinfo {address} {Amsterdam},\ \bibinfo {year} {1994})\BibitemShut
  {NoStop}%
\bibitem [{\citenamefont {Hughes}(1995)}]{hughes}%
  \BibitemOpen
  \bibfield  {author} {\bibinfo {author} {\bibfnamefont {B.}~\bibnamefont
  {Hughes}},\ }\href@noop {} {\emph {\bibinfo {title} {Random walks and random
  environments}}}\ (\bibinfo  {publisher} {Clarendon Press},\ \bibinfo
  {address} {Oxford (UK)},\ \bibinfo {year} {1995})\BibitemShut {NoStop}%
\bibitem [{\citenamefont {Lov{\'a}sz}(1996)}]{lovasz}%
  \BibitemOpen
  \bibfield  {author} {\bibinfo {author} {\bibfnamefont {L.}~\bibnamefont
  {Lov{\'a}sz}},\ }in\ \href@noop {} {\emph {\bibinfo {booktitle}
  {Combinatorics, Paul Erd{\"o}s is Eighty}}}\ (\bibinfo  {publisher}
  {J{\'a}nos Bolyai Mathematical Society, Budapest},\ \bibinfo {year} {1996})\
  p.\ \bibinfo {pages} {353}\BibitemShut {NoStop}%
\bibitem [{\citenamefont {Adamic}\ \emph {et~al.}(2001)\citenamefont {Adamic},
  \citenamefont {Lukose}, \citenamefont {Puniyani},\ and\ \citenamefont
  {Huberman}}]{PhysRevE.64.046135}%
  \BibitemOpen
  \bibfield  {author} {\bibinfo {author} {\bibfnamefont {L.~A.}\ \bibnamefont
  {Adamic}}, \bibinfo {author} {\bibfnamefont {R.~M.}\ \bibnamefont {Lukose}},
  \bibinfo {author} {\bibfnamefont {A.~R.}\ \bibnamefont {Puniyani}}, \ and\
  \bibinfo {author} {\bibfnamefont {B.~A.}\ \bibnamefont {Huberman}},\
  }\href@noop {} {\bibfield  {journal} {\bibinfo  {journal} {Phys. Rev. E}\
  }\textbf {\bibinfo {volume} {64}},\ \bibinfo {pages} {046135} (\bibinfo
  {year} {2001})}\BibitemShut {NoStop}%
\bibitem [{\citenamefont {Lv}\ \emph {et~al.}(2002)\citenamefont {Lv},
  \citenamefont {Cao}, \citenamefont {Cohen}, \citenamefont {Li},\ and\
  \citenamefont {Shenker}}]{Lv:2002}%
  \BibitemOpen
  \bibfield  {author} {\bibinfo {author} {\bibfnamefont {Q.}~\bibnamefont
  {Lv}}, \bibinfo {author} {\bibfnamefont {P.}~\bibnamefont {Cao}}, \bibinfo
  {author} {\bibfnamefont {E.}~\bibnamefont {Cohen}}, \bibinfo {author}
  {\bibfnamefont {K.}~\bibnamefont {Li}}, \ and\ \bibinfo {author}
  {\bibfnamefont {S.}~\bibnamefont {Shenker}},\ }in\ \href@noop {} {\emph
  {\bibinfo {booktitle} {Proceedings of the 16th international conference on
  Supercomputing}}}\ (\bibinfo  {publisher} {ACM Press},\ \bibinfo {address}
  {New York, NY, USA},\ \bibinfo {year} {2002})\ pp.\ \bibinfo {pages}
  {84--95}\BibitemShut {NoStop}%
\bibitem [{Soc()}]{Sociopatternsweb}%
  \BibitemOpen
  \href@noop {} {}\bibinfo {howpublished}
  {\url{http://www.sociopatterns.org/}}\BibitemShut {NoStop}%
\bibitem [{\citenamefont {Barrat}\ \emph {et~al.}(2004)\citenamefont {Barrat},
  \citenamefont {Barth{\'e}lemy}, \citenamefont {Pastor-Satorras},\ and\
  \citenamefont {Vespignani}}]{Barrat16032004}%
  \BibitemOpen
  \bibfield  {author} {\bibinfo {author} {\bibfnamefont {A.}~\bibnamefont
  {Barrat}}, \bibinfo {author} {\bibfnamefont {M.}~\bibnamefont
  {Barth{\'e}lemy}}, \bibinfo {author} {\bibfnamefont {R.}~\bibnamefont
  {Pastor-Satorras}}, \ and\ \bibinfo {author} {\bibfnamefont {A.}~\bibnamefont
  {Vespignani}},\ }\href@noop {} {\bibfield  {journal} {\bibinfo  {journal}
  {Proc. Natl. Acad. Sci. USA}\ }\textbf {\bibinfo {volume} {101}},\ \bibinfo
  {pages} {3747} (\bibinfo {year} {2004})}\BibitemShut {NoStop}%
\bibitem [{\citenamefont {Noh}\ and\ \citenamefont
  {Rieger}(2004)}]{PhysRevLett.92.118701}%
  \BibitemOpen
  \bibfield  {author} {\bibinfo {author} {\bibfnamefont {J.~D.}\ \bibnamefont
  {Noh}}\ and\ \bibinfo {author} {\bibfnamefont {H.}~\bibnamefont {Rieger}},\
  }\href {\doibase 10.1103/PhysRevLett.92.118701} {\bibfield  {journal}
  {\bibinfo  {journal} {Phys. Rev. Lett.}\ }\textbf {\bibinfo {volume} {92}},\
  \bibinfo {pages} {118701} (\bibinfo {year} {2004})}\BibitemShut {NoStop}%
\bibitem [{\citenamefont {Wu}\ \emph {et~al.}(2007)\citenamefont {Wu},
  \citenamefont {Xu}, \citenamefont {Wu},\ and\ \citenamefont
  {Wang}}]{wu07:_walks}%
  \BibitemOpen
  \bibfield  {author} {\bibinfo {author} {\bibfnamefont {A.-C.}\ \bibnamefont
  {Wu}}, \bibinfo {author} {\bibfnamefont {X.-J.}\ \bibnamefont {Xu}}, \bibinfo
  {author} {\bibfnamefont {Z.-X.}\ \bibnamefont {Wu}}, \ and\ \bibinfo {author}
  {\bibfnamefont {Y.-H.}\ \bibnamefont {Wang}},\ }\href@noop {} {\bibfield
  {journal} {\bibinfo  {journal} {Chin. Phys. Lett.}\ }\textbf {\bibinfo
  {volume} {24}},\ \bibinfo {pages} {577} (\bibinfo {year} {2007})}\BibitemShut
  {NoStop}%
\bibitem [{\citenamefont {Stauffer}\ and\ \citenamefont
  {Sahimi}(2005)}]{stauffer_annealed2005}%
  \BibitemOpen
  \bibfield  {author} {\bibinfo {author} {\bibfnamefont {D.}~\bibnamefont
  {Stauffer}}\ and\ \bibinfo {author} {\bibfnamefont {M.}~\bibnamefont
  {Sahimi}},\ }\href@noop {} {\bibfield  {journal} {\bibinfo  {journal} {Phys.
  Rev. E}\ }\textbf {\bibinfo {volume} {72}},\ \bibinfo {pages} {46128}
  (\bibinfo {year} {2005})}\BibitemShut {NoStop}%
\bibitem [{\citenamefont {Almaas}\ \emph {et~al.}(2003)\citenamefont {Almaas},
  \citenamefont {Kulkarni},\ and\ \citenamefont {Stroud}}]{almaas03:_scaling}%
  \BibitemOpen
  \bibfield  {author} {\bibinfo {author} {\bibfnamefont {E.}~\bibnamefont
  {Almaas}}, \bibinfo {author} {\bibfnamefont {R.~V.}\ \bibnamefont
  {Kulkarni}}, \ and\ \bibinfo {author} {\bibnamefont {Stroud}},\ }\href@noop
  {} {\bibfield  {journal} {\bibinfo  {journal} {Phys. Rev. E}\ }\textbf
  {\bibinfo {volume} {68}},\ \bibinfo {pages} {056105} (\bibinfo {year}
  {2003})}\BibitemShut {NoStop}%
\bibitem [{\citenamefont {Newman}(2010)}]{Newman2010}%
  \BibitemOpen
  \bibfield  {author} {\bibinfo {author} {\bibfnamefont {M.~E.~J.}\
  \bibnamefont {Newman}},\ }\href@noop {} {\emph {\bibinfo {title} {Networks:
  An introduction}}}\ (\bibinfo  {publisher} {Oxford University Press},\
  \bibinfo {address} {Oxford},\ \bibinfo {year} {2010})\BibitemShut {NoStop}%
\bibitem [{\citenamefont {Stehl\'e}\ \emph
  {et~al.}(2011{\natexlab{b}})\citenamefont {Stehl\'e}, \citenamefont {Voirin},
  \citenamefont {Barrat}, \citenamefont {Cattuto}, \citenamefont {Isella},
  \citenamefont {Pinton}, \citenamefont {Quaggiotto}, \citenamefont {Van~den
  Broeck}, \citenamefont {R\'egis}, \citenamefont {Lina},\ and\ \citenamefont
  {Vanhems}}]{Stehle:2011}%
  \BibitemOpen
  \bibfield  {author} {\bibinfo {author} {\bibfnamefont {J.}~\bibnamefont
  {Stehl\'e}}, \bibinfo {author} {\bibfnamefont {N.}~\bibnamefont {Voirin}},
  \bibinfo {author} {\bibfnamefont {A.}~\bibnamefont {Barrat}}, \bibinfo
  {author} {\bibfnamefont {C.}~\bibnamefont {Cattuto}}, \bibinfo {author}
  {\bibfnamefont {L.}~\bibnamefont {Isella}}, \bibinfo {author} {\bibfnamefont
  {J.-F.}\ \bibnamefont {Pinton}}, \bibinfo {author} {\bibfnamefont
  {M.}~\bibnamefont {Quaggiotto}}, \bibinfo {author} {\bibfnamefont
  {W.}~\bibnamefont {Van~den Broeck}}, \bibinfo {author} {\bibfnamefont
  {C.}~\bibnamefont {R\'egis}}, \bibinfo {author} {\bibfnamefont
  {B.}~\bibnamefont {Lina}}, \ and\ \bibinfo {author} {\bibfnamefont
  {P.}~\bibnamefont {Vanhems}},\ }\href {\doibase 10.1371/journal.pone.0023176}
  {\bibfield  {journal} {\bibinfo  {journal} {PLoS ONE}\ }\textbf {\bibinfo
  {volume} {6}},\ \bibinfo {pages} {e23176} (\bibinfo {year}
  {2011}{\natexlab{b}})}\BibitemShut {NoStop}%
\bibitem [{\citenamefont {Kostakos}(2009)}]{Kostakos:2009}%
  \BibitemOpen
  \bibfield  {author} {\bibinfo {author} {\bibfnamefont {V.}~\bibnamefont
  {Kostakos}},\ }\href {\doibase 10.1016/j.physa.2008.11.021} {\bibfield
  {journal} {\bibinfo  {journal} {Physica A: Statistical Mechanics and its
  Applications}\ }\textbf {\bibinfo {volume} {388}},\ \bibinfo {pages} {1007 }
  (\bibinfo {year} {2009})}\BibitemShut {NoStop}%
\bibitem [{\citenamefont {Nicosia}\ \emph {et~al.}(2011)\citenamefont
  {Nicosia}, \citenamefont {Tang}, \citenamefont {Musolesi}, \citenamefont
  {Russo}, \citenamefont {Mascolo},\ and\ \citenamefont
  {Latora}}]{Nicosia:2011}%
  \BibitemOpen
  \bibfield  {author} {\bibinfo {author} {\bibfnamefont {V.}~\bibnamefont
  {Nicosia}}, \bibinfo {author} {\bibfnamefont {J.}~\bibnamefont {Tang}},
  \bibinfo {author} {\bibfnamefont {M.}~\bibnamefont {Musolesi}}, \bibinfo
  {author} {\bibfnamefont {G.}~\bibnamefont {Russo}}, \bibinfo {author}
  {\bibfnamefont {C.}~\bibnamefont {Mascolo}}, \ and\ \bibinfo {author}
  {\bibfnamefont {V.}~\bibnamefont {Latora}},\ }\href@noop {} {\bibfield
  {journal} {\bibinfo  {journal} {arXiv:1106.2134}\ } (\bibinfo {year}
  {2011})}\BibitemShut {NoStop}%
\bibitem [{\citenamefont {Barab\'asi}(2005)}]{barabasi2005origin}%
  \BibitemOpen
  \bibfield  {author} {\bibinfo {author} {\bibfnamefont {A.}~\bibnamefont
  {Barab\'asi}},\ }\href@noop {} {\bibfield  {journal} {\bibinfo  {journal}
  {Nature}\ }\textbf {\bibinfo {volume} {435}},\ \bibinfo {pages} {207}
  (\bibinfo {year} {2005})}\BibitemShut {NoStop}%
\bibitem [{\citenamefont {den Broeck}\ \emph {et~al.}(2010)\citenamefont {den
  Broeck}, \citenamefont {Cattuto}, \citenamefont {Barrat}, \citenamefont
  {Szomsor}, \citenamefont {Correndo},\ and\ \citenamefont {Alani}}]{percol}%
  \BibitemOpen
  \bibfield  {author} {\bibinfo {author} {\bibfnamefont {W.~V.}\ \bibnamefont
  {den Broeck}}, \bibinfo {author} {\bibfnamefont {C.}~\bibnamefont {Cattuto}},
  \bibinfo {author} {\bibfnamefont {A.}~\bibnamefont {Barrat}}, \bibinfo
  {author} {\bibfnamefont {M.}~\bibnamefont {Szomsor}}, \bibinfo {author}
  {\bibfnamefont {G.}~\bibnamefont {Correndo}}, \ and\ \bibinfo {author}
  {\bibfnamefont {H.}~\bibnamefont {Alani}},\ }in\ \href@noop {} {\emph
  {\bibinfo {booktitle} {Proceedings of the 8th Annual IEEE International
  Conference on Pervasive Computing and Communications}}}\ (\bibinfo {year}
  {2010})\ p.\ \bibinfo {pages} {226}\BibitemShut {NoStop}%
\bibitem [{\citenamefont {Kossinets}\ \emph {et~al.}(2008)\citenamefont
  {Kossinets}, \citenamefont {Kleinberg},\ and\ \citenamefont
  {Watts}}]{Kleinberg:2008}%
  \BibitemOpen
  \bibfield  {author} {\bibinfo {author} {\bibfnamefont {G.}~\bibnamefont
  {Kossinets}}, \bibinfo {author} {\bibfnamefont {J.}~\bibnamefont
  {Kleinberg}}, \ and\ \bibinfo {author} {\bibfnamefont {D.}~\bibnamefont
  {Watts}},\ }in\ \href@noop {} {\emph {\bibinfo {booktitle} {Proceedings of
  the 14$^th$ ACM SIGKDD International Conference on Knowledge Discovery and
  Data Mining}}}\ (\bibinfo {year} {2008})\BibitemShut {NoStop}%
\bibitem [{\citenamefont {Pan}\ and\ \citenamefont
  {Saram\"aki}(2011)}]{Pan:2011}%
  \BibitemOpen
  \bibfield  {author} {\bibinfo {author} {\bibfnamefont {R.~K.}\ \bibnamefont
  {Pan}}\ and\ \bibinfo {author} {\bibfnamefont {J.}~\bibnamefont
  {Saram\"aki}},\ }\href {\doibase 10.1103/PhysRevE.84.016105} {\bibfield
  {journal} {\bibinfo  {journal} {Phys. Rev. E}\ }\textbf {\bibinfo {volume}
  {84}},\ \bibinfo {pages} {016105} (\bibinfo {year} {2011})}\BibitemShut
  {NoStop}%
\bibitem [{\citenamefont {Baronchelli}\ and\ \citenamefont
  {Pastor-Satorras}(2010)}]{dynam_in_weigh_networ}%
  \BibitemOpen
  \bibfield  {author} {\bibinfo {author} {\bibfnamefont {A.}~\bibnamefont
  {Baronchelli}}\ and\ \bibinfo {author} {\bibfnamefont {R.}~\bibnamefont
  {Pastor-Satorras}},\ }\href@noop {} {\bibfield  {journal} {\bibinfo
  {journal} {Phys. Rev. E}\ }\textbf {\bibinfo {volume} {82}},\ \bibinfo
  {pages} {011111} (\bibinfo {year} {2010})}\BibitemShut {NoStop}%
\bibitem [{\citenamefont {Baronchelli}\ \emph {et~al.}(2008)\citenamefont
  {Baronchelli}, \citenamefont {Catanzaro},\ and\ \citenamefont
  {Pastor-Satorras}}]{PhysRevE.78.011114}%
  \BibitemOpen
  \bibfield  {author} {\bibinfo {author} {\bibfnamefont {A.}~\bibnamefont
  {Baronchelli}}, \bibinfo {author} {\bibfnamefont {M.}~\bibnamefont
  {Catanzaro}}, \ and\ \bibinfo {author} {\bibfnamefont {R.}~\bibnamefont
  {Pastor-Satorras}},\ }\href {\doibase 10.1103/PhysRevE.78.011114} {\bibfield
  {journal} {\bibinfo  {journal} {Phys. Rev. E}\ }\textbf {\bibinfo {volume}
  {78}},\ \bibinfo {pages} {011114} (\bibinfo {year} {2008})}\BibitemShut
  {NoStop}%
\bibitem [{\citenamefont {Baronchelli}\ and\ \citenamefont
  {Loreto}(2006)}]{baronchelli2006ring}%
  \BibitemOpen
  \bibfield  {author} {\bibinfo {author} {\bibfnamefont {A.}~\bibnamefont
  {Baronchelli}}\ and\ \bibinfo {author} {\bibfnamefont {V.}~\bibnamefont
  {Loreto}},\ }\href@noop {} {\bibfield  {journal} {\bibinfo  {journal} {Phys.
  Rev. E}\ }\textbf {\bibinfo {volume} {73}},\ \bibinfo {pages} {026103}
  (\bibinfo {year} {2006})}\BibitemShut {NoStop}%
\bibitem [{\citenamefont {Pons}\ and\ \citenamefont
  {Latapy}(2005)}]{Pons:2005}%
  \BibitemOpen
  \bibfield  {author} {\bibinfo {author} {\bibfnamefont {P.}~\bibnamefont
  {Pons}}\ and\ \bibinfo {author} {\bibfnamefont {M.}~\bibnamefont {Latapy}},\
  }in\ \href@noop {} {\emph {\bibinfo {booktitle} {Proceedings of the 20th
  International Symposium on Computer and Information Sciences (ISCIS'05)}}},\
  \bibinfo {series} {Lecture Notes in Computer Science}, Vol.\ \bibinfo
  {volume} {3733}\ (\bibinfo  {publisher} {Springer},\ \bibinfo {address}
  {Istanbul, Turkey},\ \bibinfo {year} {2005})\ pp.\ \bibinfo {pages}
  {284--293}\BibitemShut {NoStop}%
\bibitem [{\citenamefont {Braha}\ and\ \citenamefont
  {Bar-Yam}(2009)}]{Braha:2009}%
  \BibitemOpen
  \bibfield  {author} {\bibinfo {author} {\bibfnamefont {D.}~\bibnamefont
  {Braha}}\ and\ \bibinfo {author} {\bibfnamefont {Y.}~\bibnamefont
  {Bar-Yam}},\ }in\ \href@noop {} {\emph {\bibinfo {booktitle} {Adaptive
  Networks}}},\ \bibinfo {series} {Understanding Complex Systems},
  Vol.~\bibinfo {volume} {51},\ \bibinfo {editor} {edited by\ \bibinfo {editor}
  {\bibfnamefont {T.}~\bibnamefont {Gross}}\ and\ \bibinfo {editor}
  {\bibfnamefont {H.}~\bibnamefont {Sayama}}}\ (\bibinfo  {publisher} {Springer
  Berlin / Heidelberg},\ \bibinfo {year} {2009})\ pp.\ \bibinfo {pages}
  {39--50}\BibitemShut {NoStop}%
\bibitem [{\citenamefont {Tang}\ \emph
  {et~al.}(2010{\natexlab{b}})\citenamefont {Tang}, \citenamefont {Musolesi},
  \citenamefont {Mascolo}, \citenamefont {Latora},\ and\ \citenamefont
  {Nicosia}}]{Tang:2010proc}%
  \BibitemOpen
  \bibfield  {author} {\bibinfo {author} {\bibfnamefont {J.}~\bibnamefont
  {Tang}}, \bibinfo {author} {\bibfnamefont {M.}~\bibnamefont {Musolesi}},
  \bibinfo {author} {\bibfnamefont {C.}~\bibnamefont {Mascolo}}, \bibinfo
  {author} {\bibfnamefont {V.}~\bibnamefont {Latora}}, \ and\ \bibinfo {author}
  {\bibfnamefont {V.}~\bibnamefont {Nicosia}},\ }in\ \href {\doibase
  http://doi.acm.org/10.1145/1852658.1852661} {\emph {\bibinfo {booktitle}
  {Proceedings of the 3rd Workshop on Social Network Systems}}},\ \bibinfo
  {series and number} {SNS '10}\ (\bibinfo  {publisher} {ACM},\ \bibinfo
  {address} {New York, NY, USA},\ \bibinfo {year} {2010})\ pp.\ \bibinfo
  {pages} {3:1--3:6}\BibitemShut {NoStop}%
\bibitem [{\citenamefont {Lerman}\ \emph {et~al.}(2010)\citenamefont {Lerman},
  \citenamefont {Ghosh},\ and\ \citenamefont {Kang}}]{Lerman:2010}%
  \BibitemOpen
  \bibfield  {author} {\bibinfo {author} {\bibfnamefont {K.}~\bibnamefont
  {Lerman}}, \bibinfo {author} {\bibfnamefont {R.}~\bibnamefont {Ghosh}}, \
  and\ \bibinfo {author} {\bibfnamefont {J.~H.}\ \bibnamefont {Kang}},\ }in\
  \href {http://doi.acm.org/10.1145/1830252.1830262} {\emph {\bibinfo
  {booktitle} {Proceedings of the Eighth Workshop on Mining and Learning with
  Graphs}}},\ \bibinfo {series and number} {MLG '10}\ (\bibinfo  {publisher}
  {ACM},\ \bibinfo {address} {New York, NY, USA},\ \bibinfo {year} {2010})\
  pp.\ \bibinfo {pages} {70--77}\BibitemShut {NoStop}%
\bibitem [{\citenamefont {Newman}(2005)}]{Newman:2005}%
  \BibitemOpen
  \bibfield  {author} {\bibinfo {author} {\bibfnamefont {M.~J.}\ \bibnamefont
  {Newman}},\ }\href@noop {} {\bibfield  {journal} {\bibinfo  {journal} {Social
  Networks}\ }\textbf {\bibinfo {volume} {27}},\ \bibinfo {pages} {39}
  (\bibinfo {year} {2005})}\BibitemShut {NoStop}%
\end{thebibliography}
\end{document}